\documentclass[prb,aps,superscriptaddress,twocolumn,notitlepage,showpacs,nofootinbib]{revtex4-1}
\usepackage{latexsym}
\usepackage{amsmath, dsfont, physics}
\usepackage{amssymb}
\usepackage{graphicx}
\usepackage{caption}
\usepackage{subfigure}
\usepackage{float}
\usepackage{mathrsfs}
\usepackage{color}
\usepackage{txfonts}
\usepackage[justification=centering,
            format=plain]{caption}

\renewcommand{\raggedright}{\leftskip=0pt \rightskip=0pt plus 0cm}

\begin{document}

\title{Nonlinear construction of topological SSH models}

\author{Jayakrishnan M. P. Nair}
\email{jayakrishnan00213@tamu.edu}
\affiliation{Institute for Quantum Science and Engineering, Texas A$\&$M University, College Station, TX 77843, USA}
\affiliation{Department of Physics and Astronomy, Texas A$\&$M University, College Station, TX 77843, USA}

\date{\today}

\begin{abstract}
The Su-Schrieffer-Heeger (SSH) model describes a paradigmatic one-dimensional (1-D) system that exhibits a non-trivial band topology. In this paper, we foreground a new scheme involving a 1-D chain of $2N+1$ Bosonic modes with nearest-neighbor interactions, in which, the topology of the system is regulated by the selective excitation of intrinsic anharmonicities. In the dispersive regime, where the characteristic detunings are considerably larger than the coupling strengths, the linearized system is topologically identical to an SSH model. This is marked by the emergence of zero-energy eigenmodes of the system, and we specifically illustrate its rich topology in a chain of size $N=6$. We consider an experimentally realizable lattice involving optical cavities which are coupled to collective systems characterized by Bosonic modes. By investigating the transmission due to a weak probe field, we provide a spectroscopic analysis of the system, demonstrating, for instance, the emergence of bistability at higher pumping rates. 
 
\end{abstract}

\maketitle

\section{Introduction}
Topological insulators have attracted a lot of attention in the recent years with wide ranging applications across physics \cite{RevModPhys.82.3045, RevModPhys.83.1057, RevModPhys.91.015006}. The concept was first investigated in the context of electronic systems \cite{RevModPhys.82.3045, RevModPhys.83.1057}, and analogous topological phenomena has been demonstrated recently in various other physical systems, including, for example, photonics \cite{lu2014topological}, cold atoms \cite{RevModPhys.83.1523}, and many more \cite{susstrunk2015observation, kane2014topological, paulose2015topological}. One of the key features of topological materials is the existence of topologically protected mid-gap states, which are robust against environmental loss and disorder. They have been realized in various platforms with frequencies ranging from microwave \cite{wang2009observation, cheng2016robust} to optical domain \cite{hafezi2013imaging, hafezi2011robust}, rendering a myriad of new avenues, specifically in the quantum domain \cite{PhysRevLett.119.023603, PhysRevLett.115.045303, PhysRevLett.118.073602, PhysRevLett.119.173901, PhysRevLett.124.023603, PhysRevLett.127.250402, barik2018topological}. Some of its applications are topological qubits \cite{kitaev2001unpaired, PhysRevB.81.014505, alicea2011non, PhysRevLett.110.076401, you2014encoding}, lasers \cite{st2017lasing, bahari2017nonreciprocal, harari2018topological, bandres2018topological} etc., all of which are key to futuristic quantum information devices. 

An archetypical model for topological materials is the Su-Schrieffer-Heeger (SSH) model for one-dimensional (1-D) condensed matter systems. A chain with staggered nearest neighbor couplings, the model was initially used to explain the properties of the organic molecule polyacetylene. Since then, several analogues of the model were realized in experiments, for example, in photonic latttices \cite{malkova2009observation}, metamaterials \cite{tan2014photonic}, waveguide arrays \cite{PhysRevLett.115.040402}, plasmonics \cite{PhysRevB.96.045417}, and electrical circuits \cite{lee2018topolectrical}, to name a few. Recently, a photonic dimer chain consisting of split-ring resonators were employed in experiments to observe the topological invariant of the SSH model \cite{PhysRevB.101.165427}. Some other relevant advances include two dimensional lattices \cite{PhysRevB.100.075437, PhysRevB.95.165109}, long-range interactions \cite{PhysRevB.89.085111, PhysRevB.99.035146}, periodically driven SSH models \cite{PhysRevA.92.023624, PhysRevA.98.013855}, and non-Hermitian topological systems \cite{RevModPhys.93.015005}. Nonlinear extensions to SSH models are also in vogue, with applications including, but not limited to, spectral tuning of the edge states \cite{PhysRevLett.121.163901} and self-induced topological transitions \cite{PhysRevB.93.155112}.

In this paper, we propose the use of Kerr nonlinearities in a 1-D chain of $N$ identically coupled Bosonic modes as a resource to engineer non-trivial topological phases. Such nonlinearities can be observed in several physical systems ranging from optical cavities \cite{boyd2020nonlinear} to magnetic systems \cite{PhysRevB.94.224410}, which has been a prime subject of interest, with exotic observable effects \cite{PhysRevLett.127.183202, PhysRevLett.126.180401, PhysRevLett.124.213604, PhysRevB.103.224401}. We place the nonlinear modes at even-indexed locations in the lattice and drive the system selectively to ensure nearly identical steady-state occupancies in the nonlinear modes. In the dispersive domain, that is, when the even-indexed modes are far detuned from the odd-indexed modes, and the coupling strengths are much less than the detunings, the linearized quantum system behaves as a 1-D SSH model. The topology of the system can be controlled externally by a laser pump, which directly gets manifested in the coupling strengths. The system also demonstrates bistability, and  for $N=6$, we consider an experimentally realizable array of optical cavities coupled to collective systems characterized by Bosonic modes to demonstrate, theoretically, the emergence of bistable edge states. Further, we use a spectroscopic analysis to elucidate the non-trivial topology by investigating the transmission to a weak probe field.

The manuscript is organized as follows. In section \ref{sec1}, we discuss the theoretical model of the 1-D chain and calculate the effective Hamiltonian in the dispersive domain, revealing the non-trivial topology. Following this, in section \ref{sec2}, we use an experimentally realizable system involving nonlinear optical cavities to investigate, numerically, the bistable response of system and employ a spectroscopic analysis to elucidate the bistability in the edge states. Finally, we conclude our results in section \ref{sec4}.

\section{Theoretical Model}\label{sec1}
We begin by considering the following generic Hamiltonian comprising of a chain of coherently coupled Bosonic oscillators $b_i$ and $a_i$ as depicted in Fig. (1).
\begin{align}
\mathcal{H}/\hbar=\sum_{i=1}^{N}\Omega_i b_i^{\dagger}b_i+\sum_{i=0}^{N}\omega_i a_i^{\dagger}a_i+U\sum_{i=1}^{N/2-1} a_{2i}^{\dagger 2}a_{2i}^2 \nonumber \\ 
+g\sum_{i=1}^{N} [b_i^{\dagger}(a_{i-1}+a_i)+h.c]+H_{d}.
\end{align} 
Here, $\Omega_i$ and $\omega_i$ characterizes the resonance frequencies of the modes $b_i$ and $a_i$ respectively, $U$ is a measure of the strength of Kerr nonlinearity in the modes $a_i$ and $g$, the strength of dispersive coupling between the modes. The Hamiltonian $H_d$ in Eq. (1) denotes the external driving on the Kerr nonlinear modes at frequency $\omega_d$, which takes the form 
\begin{align}
H_d=i\sum_{i=1}^{N/2-1}\mathcal{E}_i[a_{2i}^{\dagger}e^{-i\omega_d t}-a_{2i}e^{i\omega_d t}],
\end{align} 
where $\mathcal{E}_i=\sqrt{\frac{2\kappa_i P_d}{\hbar \omega_d}}$ signifies the Rabi frequency of external driving with a power of $P_d$ and $\kappa_i$ are the leakage rates of the $a_i$ modes. At this point, bear in mind that the aim of this section is to engineer a one dimensional SSH model between the modes $b_i$, wherein, the the topology of the system can be controlled externally by a laser pump. The following analysis can be legitimized for a chain of an arbitrary even $N$ number of modes. However, for simplicity, we restrict ourselves to a chain with $N=6$. The dynamics of the system density matrix $\rho$ is governed by the master equation \cite{agarwal2012quantum}
\begin{equation}
\frac{\dd \rho}{\dd t}=-\frac{i}{\hbar}[\mathcal{H},\rho]+\sum_{i=1}^{N}\gamma_i\mathcal{L}(b_i)\rho+\sum_{i=0}^{N}\kappa_i\mathcal{L}(a_i)\rho,
\end{equation}
where $\mathcal{L}$ is the Liouillian defined by its action $\mathcal{L}(\sigma)\rho=2\sigma\rho \sigma^{\dagger}-\sigma^{\dagger}\sigma\rho-\rho \sigma^{\dagger}\sigma$ and $\gamma_i$ are the rate of dissipation from the modes $b_i$. In the rotating frame of the drive, the mean value equations for the mode operators read
\begin{align}
\dot{b_i}=-i(\Delta^b_i-i\gamma)b_i-ig(a_{i-1}+a_{i}),\\
\dot{a_i}=-i(\Delta^a_i-i\kappa)a_i-ig\sum_{j=1}^{N}b_j(\delta_{i,j-1}+\delta_{i,j})\nonumber \\
-2iU\sum_{j=1}^{N/2-1}a_i^\dagger a_i a_i \delta_{i,2j}+\mathcal{E}\sum_{j=1}^{N/2-1}\delta_{i,2j},
\end{align}
\begin{figure}
 \captionsetup{justification=raggedright,singlelinecheck=false}
 \centering
   \includegraphics[scale=0.51]{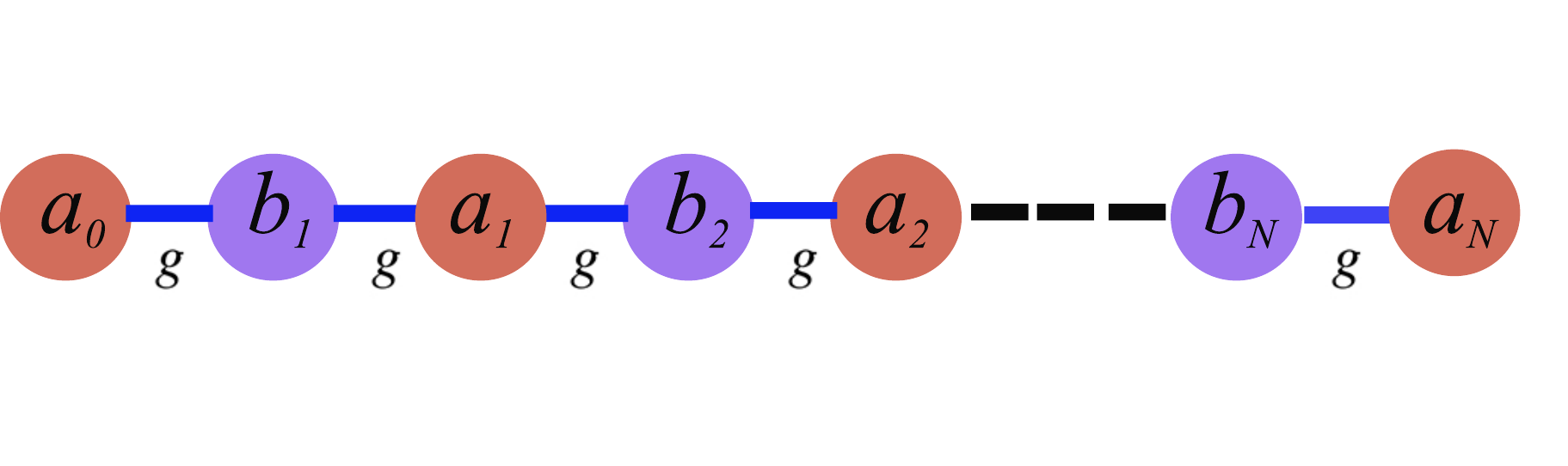}
\caption{Schematic of the coupled oscillator system described by the Hamiltonian in Eq. (1)}
\label{sch}
\end{figure}
where $\Delta^a_i=\Omega_i-\omega_d$, $\Delta^b_i=\omega_i-\omega_d$ and $\delta_{i,j}$ is the Kronecker delta function. When the system parameters fullfil the stability conditions, the mode operators converge into a constant steady state value in the long time limit, that is, $(a_i,b_i)\rightarrow(a_i^s, b_i^s)$. The steady state amplitudes can be obtained directly by solving Eq. (4,5) by setting $\dot{b_i}=\dot{a_i}=0$. Since the system is driven externally by a strong pump, the quantities $a_i^s$, $b_i^s$ satisfy $\{|a_i^s|,|b_i^s|\}\gg1$. Therefore, we can safely approximate the mode operators up to first order in quantum fluctuations as $a_i=a_i^s+\delta a_i$, $b_i=b_i^s+\delta b_i$. The effective bilinear Hamiltonian in the fluctuations is provided by
\begin{align}
\mathcal{H}_Q/\hbar=\sum_{i=1}^{N}\Delta^b_i \delta b_i^{\dagger}\delta b_i+\sum_{i=0}^{N}\tilde{\Delta}^a_i \delta a_i^{\dagger} \delta a_i\nonumber \\+\sum_{i=1}^{N/2-1}[\tilde{U}_{2i} \delta a_{2i}^{\dagger 2}+ \tilde{U}_{2i}^{*}\delta a_{2i}^2]  
+g\sum_{i=1}^{N} [\delta b_i^{\dagger}(\delta a_{i-1}+ \delta a_i)+h.c],
\end{align}
where $\tilde{\Delta}^a_i=\Delta^a_i+4\sum_{j=1}^{N/2-1}\delta_{i,2j}U|a_i^{s}|^2$, $\tilde{U}_{2i}=U(a^{s}_{2i})^2$. We define $\theta_{2i}=\text{arctan}(\frac{\text{Im}(\tilde{U}_{2i})}{\text{Re}(\tilde{U}_{2i})})$, where $\text{Re}(\tilde{U}_{2i})$, $\text{Im}(\tilde{U}_{2i})$ are the real and imaginary parts of $\tilde{U}_{2i}$ respectively. Note, \textit{en passant}, we operate in the region of parameters where the detunings $\{\Delta^b_i, \tilde{\Delta}^a_i\}$ are significantly greater than the decay parameters $\{\kappa_i, \gamma_i\}$.  Therefore, it is fair to approximate $\theta_{2i}\approx 0$. To bring the above Hamiltonian into a familiar form, we introduce Bogoliubov transformation involving the Kerr nonlinear modes $\delta\alpha_{2i}=\cosh(r_{2i})\delta a_{2i}-\sinh(r_{2i})\delta a_{2i}^{\dagger}$, where the squeezing parameters $r_{2i}$ are defined by the relation $r_{2i}={\frac{1}{4}\ln\Big(\frac{\tilde{\Delta}^a_{2i}+2|\tilde{U}_{2i}|}{\tilde{\Delta}^a_{2i}-2|\tilde{U}_{2i}|}\Big)}$, and we chose $U=-|U|$. Employing the transformations, the Eq. (6) can be recast into 
\begin{widetext}
\begin{align}
\mathcal{H}_Q/\hbar=\sum_{i=1}^{N}\frac{\Delta^b_i}{2} \delta b_i^{\dagger}\delta b_i+\sum_{i=0}^{N}\Big(1-\sum_{j=1}^{N/2-1}\delta_{i,2j}\Big)\frac{{\Delta}^a_i}{2} \delta a_i^{\dagger} \delta a_i+\sum_{i=0}^{N}\sum_{j=1}^{N/2-1}\delta_{i,2j}\frac{\xi^{a}_i}{2} \delta \alpha_i^{\dagger} \delta \alpha_i  
+g\sum_{i,k=0}^{N}\Big(1-\sum_{j=1}^{N/2-1}\delta_{i,2j}\Big)\delta a_i^{\dagger}(\delta_{k-1,i}+\delta_{k,i})b_{k}\nonumber\\
+g\sum_{i,k=0}^{N}\sum_{j=1}^{N/2-1}\delta_{i,2j}\Big(\cosh(r_{i})\delta \alpha_i+\sinh(r_{i})\delta \alpha^\dagger\Big)\Big(\delta_{k-1,i}+\delta_{k,i}\Big)b_{k}+h.c,
\end{align}
\end{widetext}
where, $\xi^{a}_{2i}=\sqrt{(\tilde{\Delta}^a_{2i})^2-4|(U{a}_{2i}^{s})^2|^2}$. The above equation describes the Hamiltonian of a set of Bosonic modes interacting via a dipole-dipole form of coupling. Note that the Bogoliubov modes $\alpha_i$ are coupled coherently with the $b_i$ modes, in which, the interaction strength is a function of the steady state occupancy of the $a_{2i}$ modes. The equations determining the steady state amplitudes are highly nonlinear in nature, which, may lead to the emergence of bistability and multistability in the coupling strengths. This will be discussed in detail in the subsequent sections. The parameters $r_{2i}$ can be significantly enhanced if we regulate the detunings ${\Delta}^a_{2i}$ close to the instability boundary, that is, ${\Delta}^a_{2i}\rightarrow 2|\tilde{U}_{2i}|$. In this limit, we have $e^{-2r_{2i}}<<1$ and $\cosh(r_{i})\approx \sinh(r_{i})\approx e^{r_{2i}}/2$. This limit in conjunction with the dispersive domain of parameters under the rotating-wave approximation (RWA) i.e., when $\{g, ge^{r_{2i}}/2\}<<\{|\Delta^b_i-\Delta^a_j|,|\Delta^b_i-\xi^{a}_j|\}<<\{|\Delta^b_i+\Delta^a_j|,|\Delta^b_i+\xi^{a}_j|\}$, we can eke out an effective interaction from Eq. (7) (Appendix A), which reads 
\begin{align}
\mathcal{H}_{eff}/\hbar=\sum_{i=1}^{N}\Delta_r b_i^{\dagger}b_i+\sum_{i=1}^{N}\sum_{j=1}^{N/2}\delta_{i,2j-1}\Big(\frac{g^2}{\Delta-\delta}b_i^\dagger b_{i+1}\Big)\nonumber\\
+\sum_{i=1}^{N}\sum_{j=1}^{N/2-1}\delta_{i,2j}\Big(\frac{g^2e^{2r}}{4(\Delta-\delta)}b_i^\dagger b_{i+1}\Big)+h.c.
\end{align}
Here, $\Delta_r=\frac{g^2(4+e^{2r})}{8(\Delta-\delta)}$ and we have set $\Delta^b_i=\Delta$, $\Delta^a_{i\neq0,N}\approx \xi_i=\delta$ and $\Delta^a_0=\Delta^a_N\approx\frac{4\delta}{e^{2r}}$. Aside from that, the system is selectively driven to ensure that the steady state occupancy of the nonlinear modes are approximately identical, that is to say, $r_{2i}\approx r$. Note, in particular for $N=6$, symmetrical driving conditions lead to $|a^s_2|^2=|a^s_4|^2$. Eq. (8) characterizes the Hamiltonian of a 1-D SSH model with staggered nearest neighbor coupling, in which, the asymmetry in the interaction strength is controlled by the parameter $r$ as a function of the external pump power.  For odd values of $i$ ranging from $1$ to $N-1$, the modes $b_i$, $b_{i+1}$ form a unit cell with intra and inter cell couplings $V=\frac{g^2}{\Delta-\delta}$ and $W=\frac{g^2e^{2r}}{4(\Delta-\delta)}$ respectively. The momentum space Hamiltonian of the system in the frame rotating at a frequency $\Delta_r$ is given by
\begin{figure}
 \captionsetup{justification=raggedright,singlelinecheck=false}
 \centering
   \includegraphics[scale=0.48]{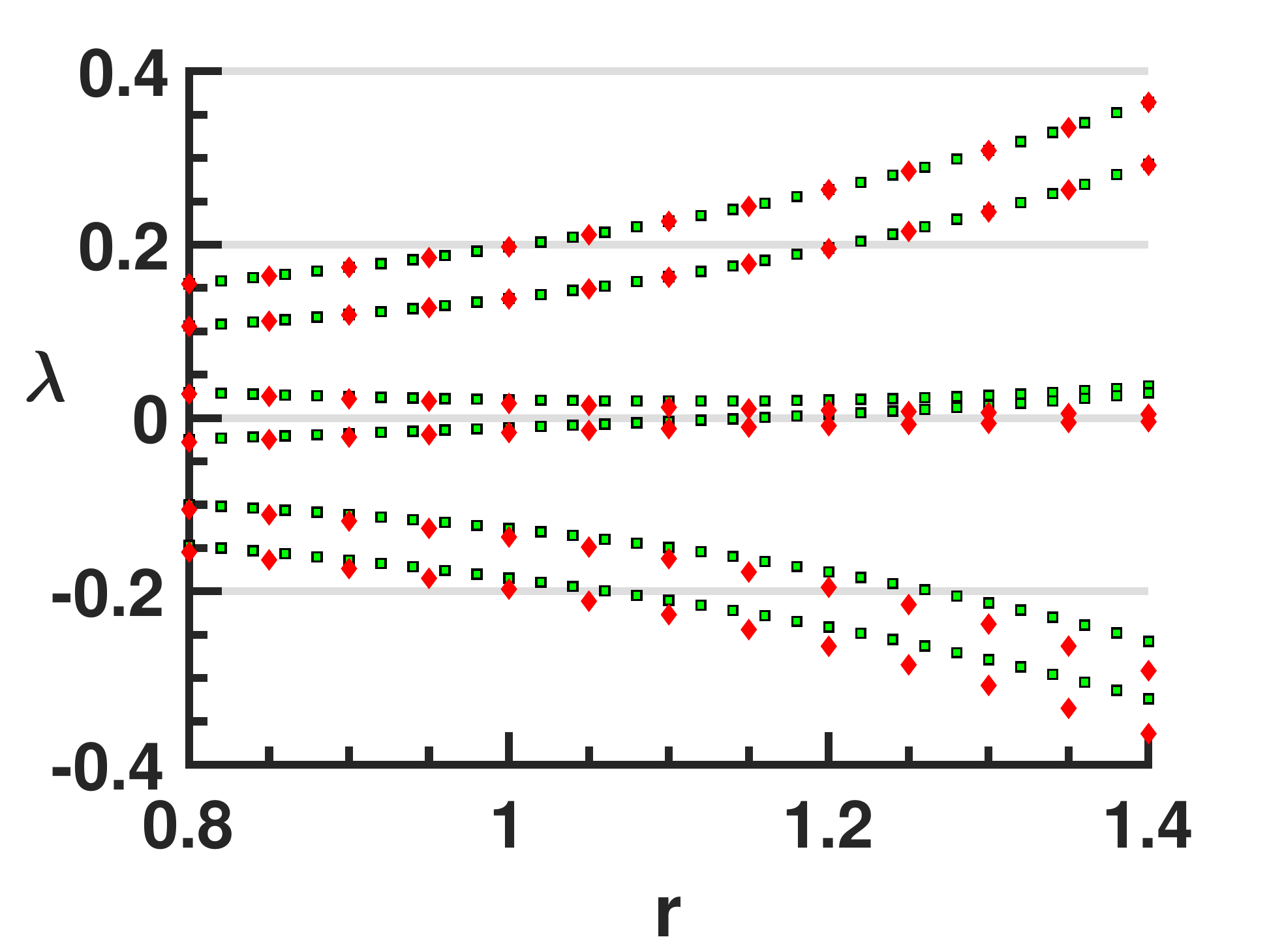}
\caption{The red diamonds and green squares represent the eigenvalues of the effective Hamiltonian in Eq. (8) and the Hamiltonian in Eq. (7) respectively for $g=1$. We have ignored the far detuned eigenvalues of $\mathcal{H}_Q$ and set the detunings $\frac{\Delta^a_i-\Delta^b_i}{g}\approx \frac{\xi^{a}_i-\Delta^b_i}{g}\approx13$. Similarly for Eq. (8), we have $\frac{\delta-\Delta}{g} \approx 13$.}
\label{sch}
\end{figure}
\begin{align}
\mathcal{H}_{k}=\begin{pmatrix}
0 & h(k) \\
h^{*}(k) & 0 
\end{pmatrix},
\end{align}
where, $k$ is the lattice constant and $h(k)=V+We^{-ik}$, an effective coupling between the two collective modes of the system. One can obtain a topological invariant of the system, namely, the winding number, defined as
\begin{align}
\nu=\frac{i}{2\pi}\oint\frac{h^{\prime}(k)}{h(k)}dk.
\end{align}
The system is said to be in a topological state with $\nu=1$ for $|W|>|V|$ and in a topologically trivial state for $|W|<|V|$ with $\nu=0$. In general, a 1-D system of oscillators with next nearest neighbor couplings can precipitate in winding numbers greater 1. When the system is in a topological state with winding number $\nu$, the bulk-edge correspondence allows for 2$\nu$ number of zero energy edge states, characterized by localized populations at the edges of the 1-D chain. In Fig. (2), we display the eigenvalues of the Hamiltonian in Eq. (7), together with that of the effective description in Eq. (8), as a function of $r$. The two sets of eigenvalues shows almost perfect semblance, a testament to our effective description of the system. Notice also the emergence of two zero energy eigenmodes, flanked on either side by the bulk modes, confirming the $\nu=1$ topology of the system. A full numerical analysis of a system with experimentally realizable parameters will be discussed in the following section.

\section{Bistability and spectroscopic detection of the edge states}\label{sec2}
In this section, we consider the model with $N=6$, involving optical cavities coupled with collective systems characterized by Bosonic operators, for example, bipartite anti-ferromagnets, collection of atoms etc., oscillating at optical frequencies. In other words, the Bosonic operators $a_i$ in Eq. (1) represent the cavity modes, while the collective systems are characterized by the $b_i$ operators. The cavity modes $a_2$ and $a_4$ are Kerr nonlinear, for instance, filled with GaAs medium having $\chi^{(3)}\approx 1.4\times10^{-18}m^2/V^2$ translating into an anharmonicity $U/2\pi\approx10^{-7}$Hz. As previously described, when the system is driven externally with the modality of driving provided in Eq. (2), symmetry around $a_3$ results in $|a^s_2|^2=|a^s_4|^2=x$. The steady state amplitude $x$ is provided by the cubic equation
\begin{align}
|\tilde{\Delta}|^2x-4U[Re(\tilde{\Delta})]x^2+4U^2x^3=\mathcal{E}^2,
\end{align}
where, $\tilde{\Delta}=(\Delta^a_2-i\kappa)-g(\chi_1+\chi_2)$, the susceptibilities $\chi_1=\frac{-g}{\Delta^b_2-\frac{g^2}{\tilde{\Delta}^a_1}}$, $\chi_2=\frac{\Delta^a_3}{2(g-\frac{\Delta^b_3\Delta^a_3}{2g})}$ and the effective detunings $\tilde{\Delta}^a_1={\Delta}^a_1-\frac{g^2}{\tilde{\Delta}^b_1}$, $\tilde{\Delta}^b_1={\Delta}^b_1-\frac{g^2}{{\Delta}^a_0}$. In Fig. 3(a), we plot the cavity response $x$ for various pump powers. As we ramp up the drive power, a sharp jump is observed in the cavity response. A similar precipitous transition is observed as we lower the drive power, this time at a different point, revealing the bistable nature of the system. Keep in mind that the steady state populations are manifested as strength of coherent coupling between $b_i$ and $b_{i+1}$ for even $i$. Therefore, bistability in the cavity response may be transferred to the edge states of the SSH model. In Fig. 3(b), we show the eigenvalues of the system by sweeping the pump power in the forward direction as depicted in Fig. 3(a), viz., for the increasing direction of the power. The figure clearly illustrates the emergence of edge and bulk states in the presence of external drive power, further corroborating our theoretical model. Note the abrupt transition from the system only possessing bulk modes to one with bulk modes flanking the central edge states, which owes its origin to the swift change in cavity response. For the parameters discussed in Fig. (3), when the pump power is less than the transition point, the steady state response is not sufficiently large enough to generate an $r$ parameter, which can produce a strong disparity in the coupling strengths. A similar transition can be observed as the pump power is varied in the opposite direction, this time at a different point. It makes for a relevant observation that the region parameters in between the two transition points exhibit bistability. For a given value of pump power in the region of bistability, we plot in Fig. 3(c-d) the position dependent population $|\psi|^2$ corresponding the two central eigenvalues, matching the two states depicted in Fig. 3(a). We have mapped the length of our 1-D chain to $13$ equidistant points between $0$ and $1$, where 0 and 1 represent the location of $a_0$ and $a_N$ respectively. State 1 displays a higher value of cavity response and thereby a squeezing parameter $r\approx0.9$, which provides a huge disparity ibetween the intra and inter cell couplings. This is manifested in the population plots with a strong amplification in the intensity close to the boundaries. In contrast, state 2 has a significantly higher population of bulk states with $r\approx0.6$. This stands as a testimony to our initial prediction that the bistability in the cavity response may be transferred to the edge states. Stated differently, the population distribution across the chain demonstrates bistability, marked by the observation of pronounced edge occupancy in the direction of decreasing drive power, whereas, the increasing direction of drive power in the bistable region is dominated by the bulk states. Note, however, that the detunings in Fig. (3) can be modulated to obtain a higher value of $r$ in the direction forward and thereby observe heightened edge effects. 

\begin{figure}
 \captionsetup{justification=raggedright,singlelinecheck=false}
 \centering
   \includegraphics[scale=0.45]{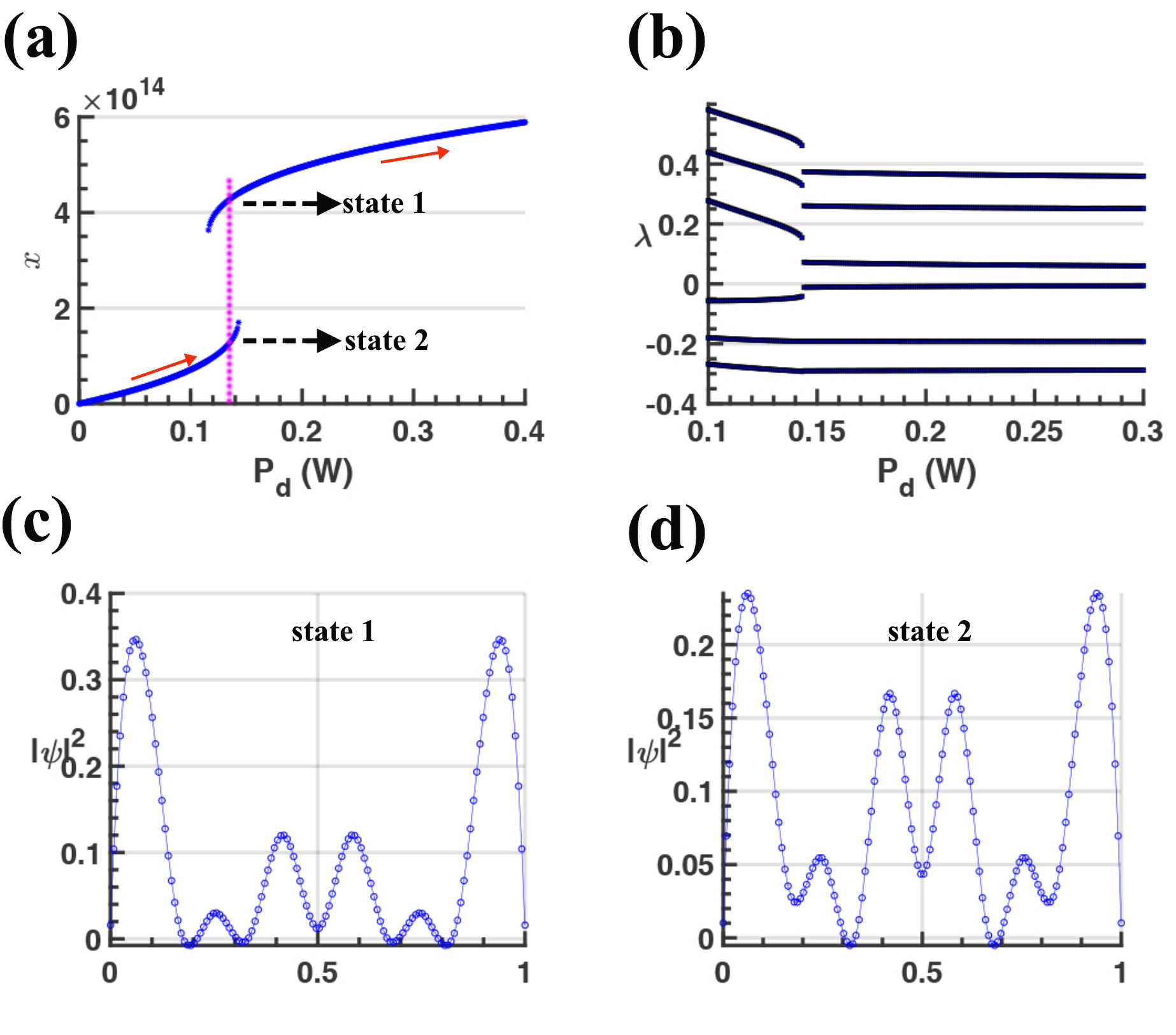}
\caption{(a) Cavity response $x$ as a function of pump power in the stable domain. The parameters $U/2\pi \approx 10^{-14}$, $\omega_d/2\pi=1.9\times10^{14}$, $\kappa_{i}/g\approx\gamma_i/g=0.01$ and $\frac{\Delta^a_i-\Delta^b_i}{g}\approx 13$; (b) Eigenvalues of the system when the drive power is scanned in the increasing direction (denoted by the red arrow in (a)); (c)-(d) probability distribution of the eigenvectors of the two central eigenvalues in (b) at $P_d=0.135W$ denoting the two distinct states in (a).}
\label{sch}
\end{figure}

The properties of our topological system, for example, the emergence of edge states can be observed in the experiments using spectroscopy, a quintessential tool and routinely applied to physical systems. The basic principle of spectroscopy is to probe the system by the application of a weak electromagnetic field, and use the transmission properties to extract key information about the system. Here, we employ a similar technique to investigate the properties of edge states using transmission spectroscopy on one of the $b_i$ modes. In the presence of a monochromatic probe field, the Hamiltonian in Eq. (7) gets modified to $\mathcal{H}_Q+\varepsilon\mathcal{H'}$, where $\mathcal{H'}=i\hbar[b_1^{\dagger}e^{-i\delta_{\text{p}}t} - h.c.]$, $\delta_{p}=\omega_{p}-\omega_{d}$, $\varepsilon=\sqrt{2\gamma \mathcal{P}_{\varepsilon}/\hbar \omega_{p}}$ and $\mathcal{P}_{\varepsilon}$ is the probe power. In the long time limit, the solution to quantum Langevin equations (QLEs) in terms of fluctuations of the mode operators may be written as 
\begin{equation}
X= \sum_{n=-\infty}^{\infty} X^{(n)} e^{-in\delta_{p} t},
\label{FE}
\end{equation}
where, $X$ belongs to $\{\delta a_0, \delta b_1,...\delta b_N, \delta a_N\}$ and $X^{0}=0$. The probe field being a weak field, we ignore the higher order terms and truncate the series at $n=\pm1$. Using the input output relations $\varepsilon+\varepsilon_r={2\gamma}b_1$ and Eq. (12), where $\varepsilon_{r}$ is the reflected field, we obtain the transmission coefficient at the probe frequency (Appendix B), 
\begin{align}
t=1+\varepsilon_{r}/\varepsilon &=2\gamma b_1^{+}/\varepsilon \notag\\
&=2\gamma(\mathcal{M}^{-1})_{22},
\end{align}
where $\mathcal{M}=i(\mathcal{H}_Q^M-\delta$) and $\mathcal{H}_Q^M$ characterizes the matrix form of $\mathcal{H}_Q$. In Fig. 4, we plot $\abs{t}$  as a function of $\delta_p$, for two different values of $P_d$. It should be noted that Fig. 4(b) at $P_d=0.135W$ portrays the state 1 in Fig. 3(a). Palpably, Fig. 4(a) is dominated by the presence of bulk modes with closely packed resonances around zero. In stark contrast to Fig. 4(a), we observe two central peaks around the origin adjoined by far detuned distinct bulk modes in Fig. 4(b), vindicating the $\nu=1$ topology of the system. 

\begin{figure}
 \captionsetup{justification=raggedright,singlelinecheck=false}
 \centering
   \includegraphics[scale=0.45]{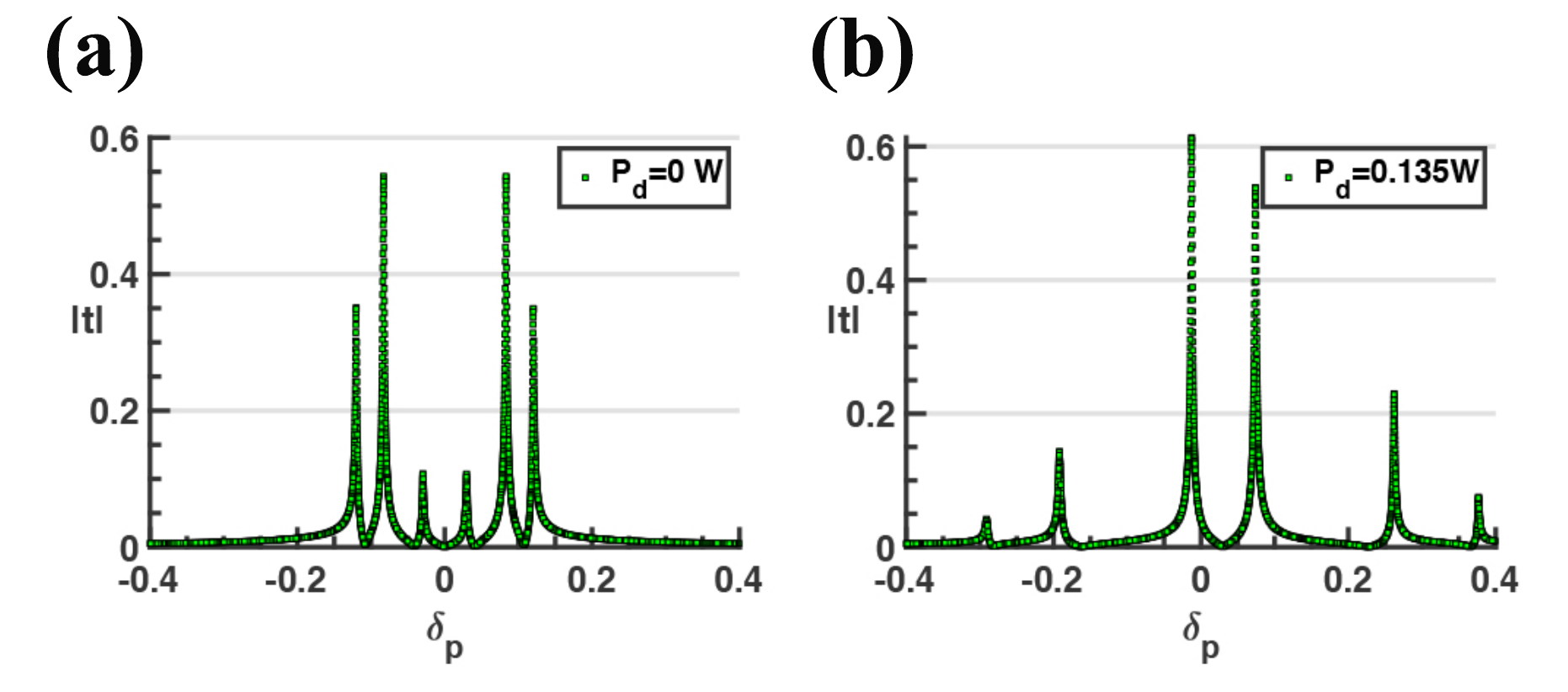}
\caption{ Absolute value of the transmission coefficient $t$ plotted as a function of $\delta_p$ for two different values of $P_d$, (a) for $P_d=0$, (b) for $P_d=0.135W$, demonstrating the spectroscopic detection of edge states. Other parameters are given in Fig. 3.}
\label{sch}
\end{figure}

\section{Summary and Concluding Remarks}\label{sec4}
In conclusion, we proposed a new scheme involving Kerr nonlinearities in a 1-D chain of $2N+1$ coupled Bosonic modes to engineer nontrivial topological phases. The topology of the system is controlled externally by a laser pump. In the absence of the pump, the system is topologically trivial, dominated by bulk modes. In contrast, the linearized Hamiltonian in the dispersive domain mimics the topology of a 1-D SSH model. The steady state occupancies of the Kerr nonlinear modes get manifested in the effective coupling mediated by the far-detuned Bogoliubov modes. For $N=6$, we HAVE illustrated the topological properties of the system in the context of an array of optical cavities coupled to collective systems, represented by Bosonic operators. For adequately large pump powers, the system goes over to the bistable domain. We obtained the probability distribution of the zero-energy modes in this domain, revealing the dominant edge-state population. Further, we used a spectroscopic analysis to theoretically illustrate the emergence of bistable edge states by examining the transmission to a weak probe field. Our model is generic, applicable to a large class of systems, including, for example, atomic ensembles, quantum dots, coupled cavity arrays and many more.
\section{Acknowledgements}
The author acknowledges the support of Herman F. Heep and Minnie Belle Heep Texas A\&M University endowed fund and thanks G. S. Agarwal and D. Mukhopadhyay for discussions, reading the manuscript and providing constructive feedback.
\appendix
\section{Effective Hamiltonian}
\label{Appendix:a}
In the limit $e^{2r}<<1$ under RWA, the Hamiltonian in Eq.(7) reduces to
\begin{widetext}
\begin{align}
\mathcal{H}_Q/\hbar=\sum_{i=1}^{N}\frac{\Delta^b_i}{2} \delta b_i^{\dagger}\delta b_i+\sum_{i=0}^{N}\Big(1-\sum_{j=1}^{N/2-1}\delta_{i}^{2j}\Big)\frac{{\Delta}^a_i}{2} \delta a_i^{\dagger} \delta a_i+\sum_{i=0}^{N}\sum_{j=1}^{N/2-1}\delta_{i}^{2j}\frac{\xi^{a}_i}{2} \delta \alpha_i^{\dagger} \delta \alpha_i  
+g\sum_{i,k=0}^{N}\Big(1-\sum_{j=1}^{N/2-1}\delta_{i}^{2j}\Big)\delta a_i^{\dagger}(\delta_{k-1}^{i}+\delta_k^{i})b_{k}\nonumber\\
+g\sum_{i,k=0}^{N}\sum_{j=1}^{N/2-1}\delta_{i}^{2j}\frac{e^{r}}{2}\delta \alpha^\dagger\Big(\delta_{k-1}^{i}+\delta_k^{i}\Big)b_{k}+h.c.
\end{align}
\end{widetext} 
Before delving into the analysis of the full system, let us begin by considering a lower dimensional system consisting of three modes given by
\begin{align}
\mathcal{H}/\hbar=\frac{\Delta^b_1}{2} b_1^\dagger b_1+\frac{\Delta^b_2}{2} b_2^\dagger b_2+\frac{\Delta^a_1}{2} a_1^\dagger a_1+ga_1^\dagger(b_1+b_2)+h.c.
\end{align}
Upon setting $\Delta^b_1=\Delta^b_2=\Delta$, $\Delta^a_1=\delta$ and moving to the frame rotating at frequency $\Delta$, we apply the unitary transformation $\mathcal{U}=e^{\bar{\lambda}X}$ on $\mathcal{H}$ to calculate $\tilde{\mathcal{H}}=\mathcal{U}\mathcal{H}\mathcal{H}^\dagger$, where, $\bar{\lambda}=\frac{g}{\Delta-\delta}$ and
\begin{align}
\mathcal{U}=\text{exp}\Big(\frac{g}{\Delta-\delta}(b_1+b_2)a^\dagger-h.c\Big).
\end{align}
Note, however,
\begin{align}
\mathcal{U}\mathcal{H}\mathcal{H}^\dagger=\mathcal{H}+\bar{\lambda}[X,\mathcal{H}]+\frac{\bar{\lambda}^2}{2}[X,[X,H]]+....
\end{align}
In the dispersive domain, wherein,  $g<<|\Delta-\delta |$, we truncate the series up to second order in $\bar{\lambda}$ and obtain 
\begin{align}
\tilde{\mathcal{H}}=(\Delta+\frac{g^2}{\Delta-\delta})(b_1^{\dagger}b_1+b_2^\dagger b_2)+\frac{g^2}{\Delta-\delta}(b_1^{\dagger}b_2+b_2^{\dagger}b_1).
\end{align}
We now extend the analysis to the full full system Eq. (A1) and the effective Hamiltonian is given by
\begin{align}
\mathcal{H}_{eff}/\hbar=\sum_{i=1}^{N}\Delta_r b_i^{\dagger}b_i+\sum_{i=1}^{N}\sum_{j=1}^{N/2}\delta_{i}^{2j-1}\Big(\frac{g^2}{\Delta-\delta}b_i^\dagger b_{i+1}\Big)\nonumber\\
+\sum_{i=1}^{N}\sum_{j=1}^{N/2-1}\delta_{i}^{2j}\Big(\frac{g^2e^{2r}}{4(\Delta-\delta)}b_i^\dagger b_{i+1}\Big)+h.c.
\end{align}
\section{Transmission to a weak probe}
In the presence of a weak probe field, the QLEs of the system described by Eq. (7) reads
\begin{align}
\dot{U}=-i\mathcal{H}_Q^M U+F_{in},
\end{align}
where, $U^T=[$$Y$ ${Y^{*}}$] and $$Y^T=\begin{bmatrix}
   \delta a_0 & \delta b_1 & \cdots & \delta b_N & \delta a_N
\end{bmatrix}.
$$
 The operator $\mathcal{H}_Q^M$ characterizes the matrix form of $\mathcal{H}_Q$,  $F_{in}=[$$Y_{in}$ ${Y_{in}^{*}}$$]^T$ and 
 $$
Y_{in}=\begin{bmatrix}
   0 &  \varepsilon e^{-i\delta t} & \cdots &0& 0
\end{bmatrix}^T.
$$
In the long time limit, the solutions to QLEs are given by
\begin{equation}
X= \sum_{n=-\infty}^{\infty} X^{(n)} e^{-in\delta_{p} t},
\label{FE}
\end{equation}
where, $X$ is an element of $\{\delta a_0, \delta b_1,...\delta b_N, \delta a_N\}$ and $X^{0}=0$. Substituting Eq. (B2) into Eq. (B1) and truncating the series at $n=\pm1$, we obtain
\begin{align}
-i\delta(U_{1}e^{-i\delta t}-U_{2}e^{i\delta t})=-i\mathcal{H}_Q^M U+F^{in},
\end{align}
where  $U_1^T=[$$U_{+}$ ${U_{-}^{*}}$], $U_2^T=[$$U_{-}$ ${U_{+}^{*}}$] and $$U_{\pm}=\begin{bmatrix}
   \delta a_0^{\pm} & \delta b_1^{\pm} & \cdots & \delta b_N^{\pm} & \delta a_N^{\pm}
\end{bmatrix}.$$
Note that we are interested in the transmission from the mode $b_1$ oscillating at the probe frequency, that is, $\delta b_1^{+}=-i(\mathcal{H}_Q^M-\delta)_{22}\varepsilon$. This, together with the input-output relations lead to Eq. (13).
\bibliography{references}

\providecommand{\noopsort}[1]{}\providecommand{\singleletter}[1]{#1}%
\begin{thebibliography}{50}%
\makeatletter
\providecommand \@ifxundefined [1]{%
 \@ifx{#1\undefined}
}%
\providecommand \@ifnum [1]{%
 \ifnum #1\expandafter \@firstoftwo
 \else \expandafter \@secondoftwo
 \fi
}%
\providecommand \@ifx [1]{%
 \ifx #1\expandafter \@firstoftwo
 \else \expandafter \@secondoftwo
 \fi
}%
\providecommand \natexlab [1]{#1}%
\providecommand \enquote  [1]{``#1''}%
\providecommand \bibnamefont  [1]{#1}%
\providecommand \bibfnamefont [1]{#1}%
\providecommand \citenamefont [1]{#1}%
\providecommand \href@noop [0]{\@secondoftwo}%
\providecommand \href [0]{\begingroup \@sanitize@url \@href}%
\providecommand \@href[1]{\@@startlink{#1}\@@href}%
\providecommand \@@href[1]{\endgroup#1\@@endlink}%
\providecommand \@sanitize@url [0]{\catcode `\\12\catcode `\$12\catcode
  `\&12\catcode `\#12\catcode `\^12\catcode `\_12\catcode `\%12\relax}%
\providecommand \@@startlink[1]{}%
\providecommand \@@endlink[0]{}%
\providecommand \url  [0]{\begingroup\@sanitize@url \@url }%
\providecommand \@url [1]{\endgroup\@href {#1}{\urlprefix }}%
\providecommand \urlprefix  [0]{URL }%
\providecommand \Eprint [0]{\href }%
\providecommand \doibase [0]{http://dx.doi.org/}%
\providecommand \selectlanguage [0]{\@gobble}%
\providecommand \bibinfo  [0]{\@secondoftwo}%
\providecommand \bibfield  [0]{\@secondoftwo}%
\providecommand \translation [1]{[#1]}%
\providecommand \BibitemOpen [0]{}%
\providecommand \bibitemStop [0]{}%
\providecommand \bibitemNoStop [0]{.\EOS\space}%
\providecommand \EOS [0]{\spacefactor3000\relax}%
\providecommand \BibitemShut  [1]{\csname bibitem#1\endcsname}%
\let\auto@bib@innerbib\@empty
\bibitem [{\citenamefont {Hasan}\ and\ \citenamefont
  {Kane}()}]{RevModPhys.82.3045}%
  \BibitemOpen
  \bibfield  {author} {\bibinfo {author} {\bibfnamefont {M.~Z.}\ \bibnamefont
  {Hasan}}\ and\ \bibinfo {author} {\bibfnamefont {C.~L.}\ \bibnamefont
  {Kane}},\ }\href@noop {} {\bibfield  {journal} {\bibinfo  {journal} {Rev.
  Mod. Phys.}\ }\textbf {\bibinfo {volume} {82}},\ \bibinfo {pages}
  {3045}}\BibitemShut {NoStop}%
\bibitem [{\citenamefont {Qi}\ and\ \citenamefont
  {Zhang}(2011)}]{RevModPhys.83.1057}%
  \BibitemOpen
  \bibfield  {author} {\bibinfo {author} {\bibfnamefont {X.-L.}\ \bibnamefont
  {Qi}}\ and\ \bibinfo {author} {\bibfnamefont {S.-C.}\ \bibnamefont {Zhang}},\
  }\href {\doibase 10.1103/RevModPhys.83.1057} {\bibfield  {journal} {\bibinfo
  {journal} {Rev. Mod. Phys.}\ }\textbf {\bibinfo {volume} {83}},\ \bibinfo
  {pages} {1057} (\bibinfo {year} {2011})}\BibitemShut {NoStop}%
\bibitem [{\citenamefont {Ozawa}\ \emph {et~al.}(2019)\citenamefont {Ozawa},
  \citenamefont {Price}, \citenamefont {Amo}, \citenamefont {Goldman},
  \citenamefont {Hafezi}, \citenamefont {Lu}, \citenamefont {Rechtsman},
  \citenamefont {Schuster}, \citenamefont {Simon}, \citenamefont {Zilberberg},\
  and\ \citenamefont {Carusotto}}]{RevModPhys.91.015006}%
  \BibitemOpen
  \bibfield  {author} {\bibinfo {author} {\bibfnamefont {T.}~\bibnamefont
  {Ozawa}}, \bibinfo {author} {\bibfnamefont {H.~M.}\ \bibnamefont {Price}},
  \bibinfo {author} {\bibfnamefont {A.}~\bibnamefont {Amo}}, \bibinfo {author}
  {\bibfnamefont {N.}~\bibnamefont {Goldman}}, \bibinfo {author} {\bibfnamefont
  {M.}~\bibnamefont {Hafezi}}, \bibinfo {author} {\bibfnamefont
  {L.}~\bibnamefont {Lu}}, \bibinfo {author} {\bibfnamefont {M.~C.}\
  \bibnamefont {Rechtsman}}, \bibinfo {author} {\bibfnamefont {D.}~\bibnamefont
  {Schuster}}, \bibinfo {author} {\bibfnamefont {J.}~\bibnamefont {Simon}},
  \bibinfo {author} {\bibfnamefont {O.}~\bibnamefont {Zilberberg}}, \ and\
  \bibinfo {author} {\bibfnamefont {I.}~\bibnamefont {Carusotto}},\ }\href
  {\doibase 10.1103/RevModPhys.91.015006} {\bibfield  {journal} {\bibinfo
  {journal} {Rev. Mod. Phys.}\ }\textbf {\bibinfo {volume} {91}},\ \bibinfo
  {pages} {015006} (\bibinfo {year} {2019})}\BibitemShut {NoStop}%
\bibitem [{\citenamefont {Lu}\ \emph {et~al.}(2014)\citenamefont {Lu},
  \citenamefont {Joannopoulos},\ and\ \citenamefont
  {Solja{\v{c}}i{\'c}}}]{lu2014topological}%
  \BibitemOpen
  \bibfield  {author} {\bibinfo {author} {\bibfnamefont {L.}~\bibnamefont
  {Lu}}, \bibinfo {author} {\bibfnamefont {J.~D.}\ \bibnamefont
  {Joannopoulos}}, \ and\ \bibinfo {author} {\bibfnamefont {M.}~\bibnamefont
  {Solja{\v{c}}i{\'c}}},\ }\href@noop {} {\bibfield  {journal} {\bibinfo
  {journal} {Nature photonics}\ }\textbf {\bibinfo {volume} {8}},\ \bibinfo
  {pages} {821} (\bibinfo {year} {2014})}\BibitemShut {NoStop}%
\bibitem [{\citenamefont {Dalibard}\ \emph {et~al.}(2011)\citenamefont
  {Dalibard}, \citenamefont {Gerbier}, \citenamefont
  {Juzeli\ifmmode~\bar{u}\else \={u}\fi{}nas},\ and\ \citenamefont
  {\"Ohberg}}]{RevModPhys.83.1523}%
  \BibitemOpen
  \bibfield  {author} {\bibinfo {author} {\bibfnamefont {J.}~\bibnamefont
  {Dalibard}}, \bibinfo {author} {\bibfnamefont {F.}~\bibnamefont {Gerbier}},
  \bibinfo {author} {\bibfnamefont {G.}~\bibnamefont
  {Juzeli\ifmmode~\bar{u}\else \={u}\fi{}nas}}, \ and\ \bibinfo {author}
  {\bibfnamefont {P.}~\bibnamefont {\"Ohberg}},\ }\href {\doibase
  10.1103/RevModPhys.83.1523} {\bibfield  {journal} {\bibinfo  {journal} {Rev.
  Mod. Phys.}\ }\textbf {\bibinfo {volume} {83}},\ \bibinfo {pages} {1523}
  (\bibinfo {year} {2011})}\BibitemShut {NoStop}%
\bibitem [{\citenamefont {S{\"u}sstrunk}\ and\ \citenamefont
  {Huber}(2015)}]{susstrunk2015observation}%
  \BibitemOpen
  \bibfield  {author} {\bibinfo {author} {\bibfnamefont {R.}~\bibnamefont
  {S{\"u}sstrunk}}\ and\ \bibinfo {author} {\bibfnamefont {S.~D.}\ \bibnamefont
  {Huber}},\ }\href@noop {} {\bibfield  {journal} {\bibinfo  {journal}
  {Science}\ }\textbf {\bibinfo {volume} {349}},\ \bibinfo {pages} {47}
  (\bibinfo {year} {2015})}\BibitemShut {NoStop}%
\bibitem [{\citenamefont {Kane}\ and\ \citenamefont
  {Lubensky}(2014)}]{kane2014topological}%
  \BibitemOpen
  \bibfield  {author} {\bibinfo {author} {\bibfnamefont {C.}~\bibnamefont
  {Kane}}\ and\ \bibinfo {author} {\bibfnamefont {T.}~\bibnamefont
  {Lubensky}},\ }\href@noop {} {\bibfield  {journal} {\bibinfo  {journal}
  {Nature Physics}\ }\textbf {\bibinfo {volume} {10}},\ \bibinfo {pages} {39}
  (\bibinfo {year} {2014})}\BibitemShut {NoStop}%
\bibitem [{\citenamefont {Paulose}\ \emph {et~al.}(2015)\citenamefont
  {Paulose}, \citenamefont {Chen},\ and\ \citenamefont
  {Vitelli}}]{paulose2015topological}%
  \BibitemOpen
  \bibfield  {author} {\bibinfo {author} {\bibfnamefont {J.}~\bibnamefont
  {Paulose}}, \bibinfo {author} {\bibfnamefont {B.~G.-g.}\ \bibnamefont
  {Chen}}, \ and\ \bibinfo {author} {\bibfnamefont {V.}~\bibnamefont
  {Vitelli}},\ }\href@noop {} {\bibfield  {journal} {\bibinfo  {journal}
  {Nature Physics}\ }\textbf {\bibinfo {volume} {11}},\ \bibinfo {pages} {153}
  (\bibinfo {year} {2015})}\BibitemShut {NoStop}%
\bibitem [{\citenamefont {Wang}\ \emph {et~al.}(2009)\citenamefont {Wang},
  \citenamefont {Chong}, \citenamefont {Joannopoulos},\ and\ \citenamefont
  {Solja{\v{c}}i{\'c}}}]{wang2009observation}%
  \BibitemOpen
  \bibfield  {author} {\bibinfo {author} {\bibfnamefont {Z.}~\bibnamefont
  {Wang}}, \bibinfo {author} {\bibfnamefont {Y.}~\bibnamefont {Chong}},
  \bibinfo {author} {\bibfnamefont {J.~D.}\ \bibnamefont {Joannopoulos}}, \
  and\ \bibinfo {author} {\bibfnamefont {M.}~\bibnamefont
  {Solja{\v{c}}i{\'c}}},\ }\href@noop {} {\bibfield  {journal} {\bibinfo
  {journal} {Nature}\ }\textbf {\bibinfo {volume} {461}},\ \bibinfo {pages}
  {772} (\bibinfo {year} {2009})}\BibitemShut {NoStop}%
\bibitem [{\citenamefont {Cheng}\ \emph {et~al.}(2016)\citenamefont {Cheng},
  \citenamefont {Jouvaud}, \citenamefont {Ni}, \citenamefont {Mousavi},
  \citenamefont {Genack},\ and\ \citenamefont {Khanikaev}}]{cheng2016robust}%
  \BibitemOpen
  \bibfield  {author} {\bibinfo {author} {\bibfnamefont {X.}~\bibnamefont
  {Cheng}}, \bibinfo {author} {\bibfnamefont {C.}~\bibnamefont {Jouvaud}},
  \bibinfo {author} {\bibfnamefont {X.}~\bibnamefont {Ni}}, \bibinfo {author}
  {\bibfnamefont {S.~H.}\ \bibnamefont {Mousavi}}, \bibinfo {author}
  {\bibfnamefont {A.~Z.}\ \bibnamefont {Genack}}, \ and\ \bibinfo {author}
  {\bibfnamefont {A.~B.}\ \bibnamefont {Khanikaev}},\ }\href@noop {} {\bibfield
   {journal} {\bibinfo  {journal} {Nature materials}\ }\textbf {\bibinfo
  {volume} {15}},\ \bibinfo {pages} {542} (\bibinfo {year} {2016})}\BibitemShut
  {NoStop}%
\bibitem [{\citenamefont {Hafezi}\ \emph {et~al.}(2013)\citenamefont {Hafezi},
  \citenamefont {Mittal}, \citenamefont {Fan}, \citenamefont {Migdall},\ and\
  \citenamefont {Taylor}}]{hafezi2013imaging}%
  \BibitemOpen
  \bibfield  {author} {\bibinfo {author} {\bibfnamefont {M.}~\bibnamefont
  {Hafezi}}, \bibinfo {author} {\bibfnamefont {S.}~\bibnamefont {Mittal}},
  \bibinfo {author} {\bibfnamefont {J.}~\bibnamefont {Fan}}, \bibinfo {author}
  {\bibfnamefont {A.}~\bibnamefont {Migdall}}, \ and\ \bibinfo {author}
  {\bibfnamefont {J.}~\bibnamefont {Taylor}},\ }\href@noop {} {\bibfield
  {journal} {\bibinfo  {journal} {Nature Photonics}\ }\textbf {\bibinfo
  {volume} {7}},\ \bibinfo {pages} {1001} (\bibinfo {year} {2013})}\BibitemShut
  {NoStop}%
\bibitem [{\citenamefont {Hafezi}\ \emph {et~al.}(2011)\citenamefont {Hafezi},
  \citenamefont {Demler}, \citenamefont {Lukin},\ and\ \citenamefont
  {Taylor}}]{hafezi2011robust}%
  \BibitemOpen
  \bibfield  {author} {\bibinfo {author} {\bibfnamefont {M.}~\bibnamefont
  {Hafezi}}, \bibinfo {author} {\bibfnamefont {E.~A.}\ \bibnamefont {Demler}},
  \bibinfo {author} {\bibfnamefont {M.~D.}\ \bibnamefont {Lukin}}, \ and\
  \bibinfo {author} {\bibfnamefont {J.~M.}\ \bibnamefont {Taylor}},\
  }\href@noop {} {\bibfield  {journal} {\bibinfo  {journal} {Nature Physics}\
  }\textbf {\bibinfo {volume} {7}},\ \bibinfo {pages} {907} (\bibinfo {year}
  {2011})}\BibitemShut {NoStop}%
\bibitem [{\citenamefont {Perczel}\ \emph {et~al.}(2017)\citenamefont
  {Perczel}, \citenamefont {Borregaard}, \citenamefont {Chang}, \citenamefont
  {Pichler}, \citenamefont {Yelin}, \citenamefont {Zoller},\ and\ \citenamefont
  {Lukin}}]{PhysRevLett.119.023603}%
  \BibitemOpen
  \bibfield  {author} {\bibinfo {author} {\bibfnamefont {J.}~\bibnamefont
  {Perczel}}, \bibinfo {author} {\bibfnamefont {J.}~\bibnamefont {Borregaard}},
  \bibinfo {author} {\bibfnamefont {D.~E.}\ \bibnamefont {Chang}}, \bibinfo
  {author} {\bibfnamefont {H.}~\bibnamefont {Pichler}}, \bibinfo {author}
  {\bibfnamefont {S.~F.}\ \bibnamefont {Yelin}}, \bibinfo {author}
  {\bibfnamefont {P.}~\bibnamefont {Zoller}}, \ and\ \bibinfo {author}
  {\bibfnamefont {M.~D.}\ \bibnamefont {Lukin}},\ }\href {\doibase
  10.1103/PhysRevLett.119.023603} {\bibfield  {journal} {\bibinfo  {journal}
  {Phys. Rev. Lett.}\ }\textbf {\bibinfo {volume} {119}},\ \bibinfo {pages}
  {023603} (\bibinfo {year} {2017})}\BibitemShut {NoStop}%
\bibitem [{\citenamefont {Pan}\ \emph {et~al.}(2015)\citenamefont {Pan},
  \citenamefont {Liu}, \citenamefont {Zhang}, \citenamefont {Yi},\ and\
  \citenamefont {Guo}}]{PhysRevLett.115.045303}%
  \BibitemOpen
  \bibfield  {author} {\bibinfo {author} {\bibfnamefont {J.-S.}\ \bibnamefont
  {Pan}}, \bibinfo {author} {\bibfnamefont {X.-J.}\ \bibnamefont {Liu}},
  \bibinfo {author} {\bibfnamefont {W.}~\bibnamefont {Zhang}}, \bibinfo
  {author} {\bibfnamefont {W.}~\bibnamefont {Yi}}, \ and\ \bibinfo {author}
  {\bibfnamefont {G.-C.}\ \bibnamefont {Guo}},\ }\href {\doibase
  10.1103/PhysRevLett.115.045303} {\bibfield  {journal} {\bibinfo  {journal}
  {Phys. Rev. Lett.}\ }\textbf {\bibinfo {volume} {115}},\ \bibinfo {pages}
  {045303} (\bibinfo {year} {2015})}\BibitemShut {NoStop}%
\bibitem [{\citenamefont {Mivehvar}\ \emph {et~al.}(2017)\citenamefont
  {Mivehvar}, \citenamefont {Ritsch},\ and\ \citenamefont
  {Piazza}}]{PhysRevLett.118.073602}%
  \BibitemOpen
  \bibfield  {author} {\bibinfo {author} {\bibfnamefont {F.}~\bibnamefont
  {Mivehvar}}, \bibinfo {author} {\bibfnamefont {H.}~\bibnamefont {Ritsch}}, \
  and\ \bibinfo {author} {\bibfnamefont {F.}~\bibnamefont {Piazza}},\ }\href
  {\doibase 10.1103/PhysRevLett.118.073602} {\bibfield  {journal} {\bibinfo
  {journal} {Phys. Rev. Lett.}\ }\textbf {\bibinfo {volume} {118}},\ \bibinfo
  {pages} {073602} (\bibinfo {year} {2017})}\BibitemShut {NoStop}%
\bibitem [{\citenamefont {Doyeux}\ \emph {et~al.}(2017)\citenamefont {Doyeux},
  \citenamefont {Gangaraj}, \citenamefont {Hanson},\ and\ \citenamefont
  {Antezza}}]{PhysRevLett.119.173901}%
  \BibitemOpen
  \bibfield  {author} {\bibinfo {author} {\bibfnamefont {P.}~\bibnamefont
  {Doyeux}}, \bibinfo {author} {\bibfnamefont {S.~A.~H.}\ \bibnamefont
  {Gangaraj}}, \bibinfo {author} {\bibfnamefont {G.~W.}\ \bibnamefont
  {Hanson}}, \ and\ \bibinfo {author} {\bibfnamefont {M.}~\bibnamefont
  {Antezza}},\ }\href {\doibase 10.1103/PhysRevLett.119.173901} {\bibfield
  {journal} {\bibinfo  {journal} {Phys. Rev. Lett.}\ }\textbf {\bibinfo
  {volume} {119}},\ \bibinfo {pages} {173901} (\bibinfo {year}
  {2017})}\BibitemShut {NoStop}%
\bibitem [{\citenamefont {Nie}\ \emph {et~al.}(2020)\citenamefont {Nie},
  \citenamefont {Peng}, \citenamefont {Nori},\ and\ \citenamefont
  {Liu}}]{PhysRevLett.124.023603}%
  \BibitemOpen
  \bibfield  {author} {\bibinfo {author} {\bibfnamefont {W.}~\bibnamefont
  {Nie}}, \bibinfo {author} {\bibfnamefont {Z.~H.}\ \bibnamefont {Peng}},
  \bibinfo {author} {\bibfnamefont {F.}~\bibnamefont {Nori}}, \ and\ \bibinfo
  {author} {\bibfnamefont {Y.-x.}\ \bibnamefont {Liu}},\ }\href {\doibase
  10.1103/PhysRevLett.124.023603} {\bibfield  {journal} {\bibinfo  {journal}
  {Phys. Rev. Lett.}\ }\textbf {\bibinfo {volume} {124}},\ \bibinfo {pages}
  {023603} (\bibinfo {year} {2020})}\BibitemShut {NoStop}%
\bibitem [{\citenamefont {Nie}\ \emph {et~al.}(2021)\citenamefont {Nie},
  \citenamefont {Antezza}, \citenamefont {Liu},\ and\ \citenamefont
  {Nori}}]{PhysRevLett.127.250402}%
  \BibitemOpen
  \bibfield  {author} {\bibinfo {author} {\bibfnamefont {W.}~\bibnamefont
  {Nie}}, \bibinfo {author} {\bibfnamefont {M.}~\bibnamefont {Antezza}},
  \bibinfo {author} {\bibfnamefont {Y.-x.}\ \bibnamefont {Liu}}, \ and\
  \bibinfo {author} {\bibfnamefont {F.}~\bibnamefont {Nori}},\ }\href {\doibase
  10.1103/PhysRevLett.127.250402} {\bibfield  {journal} {\bibinfo  {journal}
  {Phys. Rev. Lett.}\ }\textbf {\bibinfo {volume} {127}},\ \bibinfo {pages}
  {250402} (\bibinfo {year} {2021})}\BibitemShut {NoStop}%
\bibitem [{\citenamefont {Barik}\ \emph {et~al.}(2018)\citenamefont {Barik},
  \citenamefont {Karasahin}, \citenamefont {Flower}, \citenamefont {Cai},
  \citenamefont {Miyake}, \citenamefont {DeGottardi}, \citenamefont {Hafezi},\
  and\ \citenamefont {Waks}}]{barik2018topological}%
  \BibitemOpen
  \bibfield  {author} {\bibinfo {author} {\bibfnamefont {S.}~\bibnamefont
  {Barik}}, \bibinfo {author} {\bibfnamefont {A.}~\bibnamefont {Karasahin}},
  \bibinfo {author} {\bibfnamefont {C.}~\bibnamefont {Flower}}, \bibinfo
  {author} {\bibfnamefont {T.}~\bibnamefont {Cai}}, \bibinfo {author}
  {\bibfnamefont {H.}~\bibnamefont {Miyake}}, \bibinfo {author} {\bibfnamefont
  {W.}~\bibnamefont {DeGottardi}}, \bibinfo {author} {\bibfnamefont
  {M.}~\bibnamefont {Hafezi}}, \ and\ \bibinfo {author} {\bibfnamefont
  {E.}~\bibnamefont {Waks}},\ }\href@noop {} {\bibfield  {journal} {\bibinfo
  {journal} {Science}\ }\textbf {\bibinfo {volume} {359}},\ \bibinfo {pages}
  {666} (\bibinfo {year} {2018})}\BibitemShut {NoStop}%
\bibitem [{\citenamefont {Kitaev}(2001)}]{kitaev2001unpaired}%
  \BibitemOpen
  \bibfield  {author} {\bibinfo {author} {\bibfnamefont {A.~Y.}\ \bibnamefont
  {Kitaev}},\ }\href@noop {} {\bibfield  {journal} {\bibinfo  {journal}
  {Physics-uspekhi}\ }\textbf {\bibinfo {volume} {44}},\ \bibinfo {pages} {131}
  (\bibinfo {year} {2001})}\BibitemShut {NoStop}%
\bibitem [{\citenamefont {You}\ \emph {et~al.}(2010)\citenamefont {You},
  \citenamefont {Shi}, \citenamefont {Hu},\ and\ \citenamefont
  {Nori}}]{PhysRevB.81.014505}%
  \BibitemOpen
  \bibfield  {author} {\bibinfo {author} {\bibfnamefont {J.~Q.}\ \bibnamefont
  {You}}, \bibinfo {author} {\bibfnamefont {X.-F.}\ \bibnamefont {Shi}},
  \bibinfo {author} {\bibfnamefont {X.}~\bibnamefont {Hu}}, \ and\ \bibinfo
  {author} {\bibfnamefont {F.}~\bibnamefont {Nori}},\ }\href {\doibase
  10.1103/PhysRevB.81.014505} {\bibfield  {journal} {\bibinfo  {journal} {Phys.
  Rev. B}\ }\textbf {\bibinfo {volume} {81}},\ \bibinfo {pages} {014505}
  (\bibinfo {year} {2010})}\BibitemShut {NoStop}%
\bibitem [{\citenamefont {Alicea}\ \emph {et~al.}(2011)\citenamefont {Alicea},
  \citenamefont {Oreg}, \citenamefont {Refael}, \citenamefont {Von~Oppen},\
  and\ \citenamefont {Fisher}}]{alicea2011non}%
  \BibitemOpen
  \bibfield  {author} {\bibinfo {author} {\bibfnamefont {J.}~\bibnamefont
  {Alicea}}, \bibinfo {author} {\bibfnamefont {Y.}~\bibnamefont {Oreg}},
  \bibinfo {author} {\bibfnamefont {G.}~\bibnamefont {Refael}}, \bibinfo
  {author} {\bibfnamefont {F.}~\bibnamefont {Von~Oppen}}, \ and\ \bibinfo
  {author} {\bibfnamefont {M.}~\bibnamefont {Fisher}},\ }\href@noop {}
  {\bibfield  {journal} {\bibinfo  {journal} {Nature Physics}\ }\textbf
  {\bibinfo {volume} {7}},\ \bibinfo {pages} {412} (\bibinfo {year}
  {2011})}\BibitemShut {NoStop}%
\bibitem [{\citenamefont {Liu}\ \emph {et~al.}(2013)\citenamefont {Liu},
  \citenamefont {Liu},\ and\ \citenamefont {Cheng}}]{PhysRevLett.110.076401}%
  \BibitemOpen
  \bibfield  {author} {\bibinfo {author} {\bibfnamefont {X.-J.}\ \bibnamefont
  {Liu}}, \bibinfo {author} {\bibfnamefont {Z.-X.}\ \bibnamefont {Liu}}, \ and\
  \bibinfo {author} {\bibfnamefont {M.}~\bibnamefont {Cheng}},\ }\href
  {\doibase 10.1103/PhysRevLett.110.076401} {\bibfield  {journal} {\bibinfo
  {journal} {Phys. Rev. Lett.}\ }\textbf {\bibinfo {volume} {110}},\ \bibinfo
  {pages} {076401} (\bibinfo {year} {2013})}\BibitemShut {NoStop}%
\bibitem [{\citenamefont {You}\ \emph {et~al.}(2014)\citenamefont {You},
  \citenamefont {Wang}, \citenamefont {Zhang},\ and\ \citenamefont
  {Nori}}]{you2014encoding}%
  \BibitemOpen
  \bibfield  {author} {\bibinfo {author} {\bibfnamefont {J.}~\bibnamefont
  {You}}, \bibinfo {author} {\bibfnamefont {Z.}~\bibnamefont {Wang}}, \bibinfo
  {author} {\bibfnamefont {W.}~\bibnamefont {Zhang}}, \ and\ \bibinfo {author}
  {\bibfnamefont {F.}~\bibnamefont {Nori}},\ }\href@noop {} {\bibfield
  {journal} {\bibinfo  {journal} {Scientific reports}\ }\textbf {\bibinfo
  {volume} {4}},\ \bibinfo {pages} {1} (\bibinfo {year} {2014})}\BibitemShut
  {NoStop}%
\bibitem [{\citenamefont {St-Jean}\ \emph {et~al.}(2017)\citenamefont
  {St-Jean}, \citenamefont {Goblot}, \citenamefont {Galopin}, \citenamefont
  {Lema{\^\i}tre}, \citenamefont {Ozawa}, \citenamefont {Le~Gratiet},
  \citenamefont {Sagnes}, \citenamefont {Bloch},\ and\ \citenamefont
  {Amo}}]{st2017lasing}%
  \BibitemOpen
  \bibfield  {author} {\bibinfo {author} {\bibfnamefont {P.}~\bibnamefont
  {St-Jean}}, \bibinfo {author} {\bibfnamefont {V.}~\bibnamefont {Goblot}},
  \bibinfo {author} {\bibfnamefont {E.}~\bibnamefont {Galopin}}, \bibinfo
  {author} {\bibfnamefont {A.}~\bibnamefont {Lema{\^\i}tre}}, \bibinfo {author}
  {\bibfnamefont {T.}~\bibnamefont {Ozawa}}, \bibinfo {author} {\bibfnamefont
  {L.}~\bibnamefont {Le~Gratiet}}, \bibinfo {author} {\bibfnamefont
  {I.}~\bibnamefont {Sagnes}}, \bibinfo {author} {\bibfnamefont
  {J.}~\bibnamefont {Bloch}}, \ and\ \bibinfo {author} {\bibfnamefont
  {A.}~\bibnamefont {Amo}},\ }\href@noop {} {\bibfield  {journal} {\bibinfo
  {journal} {Nature Photonics}\ }\textbf {\bibinfo {volume} {11}},\ \bibinfo
  {pages} {651} (\bibinfo {year} {2017})}\BibitemShut {NoStop}%
\bibitem [{\citenamefont {Bahari}\ \emph {et~al.}(2017)\citenamefont {Bahari},
  \citenamefont {Ndao}, \citenamefont {Vallini}, \citenamefont {El~Amili},
  \citenamefont {Fainman},\ and\ \citenamefont
  {Kant{\'e}}}]{bahari2017nonreciprocal}%
  \BibitemOpen
  \bibfield  {author} {\bibinfo {author} {\bibfnamefont {B.}~\bibnamefont
  {Bahari}}, \bibinfo {author} {\bibfnamefont {A.}~\bibnamefont {Ndao}},
  \bibinfo {author} {\bibfnamefont {F.}~\bibnamefont {Vallini}}, \bibinfo
  {author} {\bibfnamefont {A.}~\bibnamefont {El~Amili}}, \bibinfo {author}
  {\bibfnamefont {Y.}~\bibnamefont {Fainman}}, \ and\ \bibinfo {author}
  {\bibfnamefont {B.}~\bibnamefont {Kant{\'e}}},\ }\href@noop {} {\bibfield
  {journal} {\bibinfo  {journal} {Science}\ }\textbf {\bibinfo {volume}
  {358}},\ \bibinfo {pages} {636} (\bibinfo {year} {2017})}\BibitemShut
  {NoStop}%
\bibitem [{\citenamefont {Harari}\ \emph {et~al.}(2018)\citenamefont {Harari},
  \citenamefont {Bandres}, \citenamefont {Lumer}, \citenamefont {Rechtsman},
  \citenamefont {Chong}, \citenamefont {Khajavikhan}, \citenamefont
  {Christodoulides},\ and\ \citenamefont {Segev}}]{harari2018topological}%
  \BibitemOpen
  \bibfield  {author} {\bibinfo {author} {\bibfnamefont {G.}~\bibnamefont
  {Harari}}, \bibinfo {author} {\bibfnamefont {M.~A.}\ \bibnamefont {Bandres}},
  \bibinfo {author} {\bibfnamefont {Y.}~\bibnamefont {Lumer}}, \bibinfo
  {author} {\bibfnamefont {M.~C.}\ \bibnamefont {Rechtsman}}, \bibinfo {author}
  {\bibfnamefont {Y.~D.}\ \bibnamefont {Chong}}, \bibinfo {author}
  {\bibfnamefont {M.}~\bibnamefont {Khajavikhan}}, \bibinfo {author}
  {\bibfnamefont {D.~N.}\ \bibnamefont {Christodoulides}}, \ and\ \bibinfo
  {author} {\bibfnamefont {M.}~\bibnamefont {Segev}},\ }\href@noop {}
  {\bibfield  {journal} {\bibinfo  {journal} {Science}\ }\textbf {\bibinfo
  {volume} {359}},\ \bibinfo {pages} {eaar4003} (\bibinfo {year}
  {2018})}\BibitemShut {NoStop}%
\bibitem [{\citenamefont {Bandres}\ \emph {et~al.}(2018)\citenamefont
  {Bandres}, \citenamefont {Wittek}, \citenamefont {Harari}, \citenamefont
  {Parto}, \citenamefont {Ren}, \citenamefont {Segev}, \citenamefont
  {Christodoulides},\ and\ \citenamefont
  {Khajavikhan}}]{bandres2018topological}%
  \BibitemOpen
  \bibfield  {author} {\bibinfo {author} {\bibfnamefont {M.~A.}\ \bibnamefont
  {Bandres}}, \bibinfo {author} {\bibfnamefont {S.}~\bibnamefont {Wittek}},
  \bibinfo {author} {\bibfnamefont {G.}~\bibnamefont {Harari}}, \bibinfo
  {author} {\bibfnamefont {M.}~\bibnamefont {Parto}}, \bibinfo {author}
  {\bibfnamefont {J.}~\bibnamefont {Ren}}, \bibinfo {author} {\bibfnamefont
  {M.}~\bibnamefont {Segev}}, \bibinfo {author} {\bibfnamefont {D.~N.}\
  \bibnamefont {Christodoulides}}, \ and\ \bibinfo {author} {\bibfnamefont
  {M.}~\bibnamefont {Khajavikhan}},\ }\href@noop {} {\bibfield  {journal}
  {\bibinfo  {journal} {Science}\ }\textbf {\bibinfo {volume} {359}},\ \bibinfo
  {pages} {eaar4005} (\bibinfo {year} {2018})}\BibitemShut {NoStop}%
\bibitem [{\citenamefont {Malkova}\ \emph {et~al.}(2009)\citenamefont
  {Malkova}, \citenamefont {Hromada}, \citenamefont {Wang}, \citenamefont
  {Bryant},\ and\ \citenamefont {Chen}}]{malkova2009observation}%
  \BibitemOpen
  \bibfield  {author} {\bibinfo {author} {\bibfnamefont {N.}~\bibnamefont
  {Malkova}}, \bibinfo {author} {\bibfnamefont {I.}~\bibnamefont {Hromada}},
  \bibinfo {author} {\bibfnamefont {X.}~\bibnamefont {Wang}}, \bibinfo {author}
  {\bibfnamefont {G.}~\bibnamefont {Bryant}}, \ and\ \bibinfo {author}
  {\bibfnamefont {Z.}~\bibnamefont {Chen}},\ }\href@noop {} {\bibfield
  {journal} {\bibinfo  {journal} {Optics letters}\ }\textbf {\bibinfo {volume}
  {34}},\ \bibinfo {pages} {1633} (\bibinfo {year} {2009})}\BibitemShut
  {NoStop}%
\bibitem [{\citenamefont {Tan}\ \emph {et~al.}(2014)\citenamefont {Tan},
  \citenamefont {Sun}, \citenamefont {Chen},\ and\ \citenamefont
  {Shen}}]{tan2014photonic}%
  \BibitemOpen
  \bibfield  {author} {\bibinfo {author} {\bibfnamefont {W.}~\bibnamefont
  {Tan}}, \bibinfo {author} {\bibfnamefont {Y.}~\bibnamefont {Sun}}, \bibinfo
  {author} {\bibfnamefont {H.}~\bibnamefont {Chen}}, \ and\ \bibinfo {author}
  {\bibfnamefont {S.-Q.}\ \bibnamefont {Shen}},\ }\href@noop {} {\bibfield
  {journal} {\bibinfo  {journal} {Scientific reports}\ }\textbf {\bibinfo
  {volume} {4}},\ \bibinfo {pages} {1} (\bibinfo {year} {2014})}\BibitemShut
  {NoStop}%
\bibitem [{\citenamefont {Zeuner}\ \emph {et~al.}(2015)\citenamefont {Zeuner},
  \citenamefont {Rechtsman}, \citenamefont {Plotnik}, \citenamefont {Lumer},
  \citenamefont {Nolte}, \citenamefont {Rudner}, \citenamefont {Segev},\ and\
  \citenamefont {Szameit}}]{PhysRevLett.115.040402}%
  \BibitemOpen
  \bibfield  {author} {\bibinfo {author} {\bibfnamefont {J.~M.}\ \bibnamefont
  {Zeuner}}, \bibinfo {author} {\bibfnamefont {M.~C.}\ \bibnamefont
  {Rechtsman}}, \bibinfo {author} {\bibfnamefont {Y.}~\bibnamefont {Plotnik}},
  \bibinfo {author} {\bibfnamefont {Y.}~\bibnamefont {Lumer}}, \bibinfo
  {author} {\bibfnamefont {S.}~\bibnamefont {Nolte}}, \bibinfo {author}
  {\bibfnamefont {M.~S.}\ \bibnamefont {Rudner}}, \bibinfo {author}
  {\bibfnamefont {M.}~\bibnamefont {Segev}}, \ and\ \bibinfo {author}
  {\bibfnamefont {A.}~\bibnamefont {Szameit}},\ }\href {\doibase
  10.1103/PhysRevLett.115.040402} {\bibfield  {journal} {\bibinfo  {journal}
  {Phys. Rev. Lett.}\ }\textbf {\bibinfo {volume} {115}},\ \bibinfo {pages}
  {040402} (\bibinfo {year} {2015})}\BibitemShut {NoStop}%
\bibitem [{\citenamefont {Bleckmann}\ \emph {et~al.}(2017)\citenamefont
  {Bleckmann}, \citenamefont {Cherpakova}, \citenamefont {Linden},\ and\
  \citenamefont {Alberti}}]{PhysRevB.96.045417}%
  \BibitemOpen
  \bibfield  {author} {\bibinfo {author} {\bibfnamefont {F.}~\bibnamefont
  {Bleckmann}}, \bibinfo {author} {\bibfnamefont {Z.}~\bibnamefont
  {Cherpakova}}, \bibinfo {author} {\bibfnamefont {S.}~\bibnamefont {Linden}},
  \ and\ \bibinfo {author} {\bibfnamefont {A.}~\bibnamefont {Alberti}},\ }\href
  {\doibase 10.1103/PhysRevB.96.045417} {\bibfield  {journal} {\bibinfo
  {journal} {Phys. Rev. B}\ }\textbf {\bibinfo {volume} {96}},\ \bibinfo
  {pages} {045417} (\bibinfo {year} {2017})}\BibitemShut {NoStop}%
\bibitem [{\citenamefont {Lee}\ \emph {et~al.}(2018)\citenamefont {Lee},
  \citenamefont {Imhof}, \citenamefont {Berger}, \citenamefont {Bayer},
  \citenamefont {Brehm}, \citenamefont {Molenkamp}, \citenamefont {Kiessling},\
  and\ \citenamefont {Thomale}}]{lee2018topolectrical}%
  \BibitemOpen
  \bibfield  {author} {\bibinfo {author} {\bibfnamefont {C.~H.}\ \bibnamefont
  {Lee}}, \bibinfo {author} {\bibfnamefont {S.}~\bibnamefont {Imhof}}, \bibinfo
  {author} {\bibfnamefont {C.}~\bibnamefont {Berger}}, \bibinfo {author}
  {\bibfnamefont {F.}~\bibnamefont {Bayer}}, \bibinfo {author} {\bibfnamefont
  {J.}~\bibnamefont {Brehm}}, \bibinfo {author} {\bibfnamefont {L.~W.}\
  \bibnamefont {Molenkamp}}, \bibinfo {author} {\bibfnamefont {T.}~\bibnamefont
  {Kiessling}}, \ and\ \bibinfo {author} {\bibfnamefont {R.}~\bibnamefont
  {Thomale}},\ }\href@noop {} {\bibfield  {journal} {\bibinfo  {journal}
  {Communications Physics}\ }\textbf {\bibinfo {volume} {1}},\ \bibinfo {pages}
  {1} (\bibinfo {year} {2018})}\BibitemShut {NoStop}%
\bibitem [{\citenamefont {Jiang}\ \emph {et~al.}(2020)\citenamefont {Jiang},
  \citenamefont {Ren}, \citenamefont {Guo}, \citenamefont {Zhu}, \citenamefont
  {Long}, \citenamefont {Jiang},\ and\ \citenamefont
  {Chen}}]{PhysRevB.101.165427}%
  \BibitemOpen
  \bibfield  {author} {\bibinfo {author} {\bibfnamefont {J.}~\bibnamefont
  {Jiang}}, \bibinfo {author} {\bibfnamefont {J.}~\bibnamefont {Ren}}, \bibinfo
  {author} {\bibfnamefont {Z.}~\bibnamefont {Guo}}, \bibinfo {author}
  {\bibfnamefont {W.}~\bibnamefont {Zhu}}, \bibinfo {author} {\bibfnamefont
  {Y.}~\bibnamefont {Long}}, \bibinfo {author} {\bibfnamefont {H.}~\bibnamefont
  {Jiang}}, \ and\ \bibinfo {author} {\bibfnamefont {H.}~\bibnamefont {Chen}},\
  }\href {\doibase 10.1103/PhysRevB.101.165427} {\bibfield  {journal} {\bibinfo
   {journal} {Phys. Rev. B}\ }\textbf {\bibinfo {volume} {101}},\ \bibinfo
  {pages} {165427} (\bibinfo {year} {2020})}\BibitemShut {NoStop}%
\bibitem [{\citenamefont {Obana}\ \emph {et~al.}(2019)\citenamefont {Obana},
  \citenamefont {Liu},\ and\ \citenamefont
  {Wakabayashi}}]{PhysRevB.100.075437}%
  \BibitemOpen
  \bibfield  {author} {\bibinfo {author} {\bibfnamefont {D.}~\bibnamefont
  {Obana}}, \bibinfo {author} {\bibfnamefont {F.}~\bibnamefont {Liu}}, \ and\
  \bibinfo {author} {\bibfnamefont {K.}~\bibnamefont {Wakabayashi}},\ }\href
  {\doibase 10.1103/PhysRevB.100.075437} {\bibfield  {journal} {\bibinfo
  {journal} {Phys. Rev. B}\ }\textbf {\bibinfo {volume} {100}},\ \bibinfo
  {pages} {075437} (\bibinfo {year} {2019})}\BibitemShut {NoStop}%
\bibitem [{\citenamefont {Arkinstall}\ \emph {et~al.}(2017)\citenamefont
  {Arkinstall}, \citenamefont {Teimourpour}, \citenamefont {Feng},
  \citenamefont {El-Ganainy},\ and\ \citenamefont
  {Schomerus}}]{PhysRevB.95.165109}%
  \BibitemOpen
  \bibfield  {author} {\bibinfo {author} {\bibfnamefont {J.}~\bibnamefont
  {Arkinstall}}, \bibinfo {author} {\bibfnamefont {M.~H.}\ \bibnamefont
  {Teimourpour}}, \bibinfo {author} {\bibfnamefont {L.}~\bibnamefont {Feng}},
  \bibinfo {author} {\bibfnamefont {R.}~\bibnamefont {El-Ganainy}}, \ and\
  \bibinfo {author} {\bibfnamefont {H.}~\bibnamefont {Schomerus}},\ }\href
  {\doibase 10.1103/PhysRevB.95.165109} {\bibfield  {journal} {\bibinfo
  {journal} {Phys. Rev. B}\ }\textbf {\bibinfo {volume} {95}},\ \bibinfo
  {pages} {165109} (\bibinfo {year} {2017})}\BibitemShut {NoStop}%
\bibitem [{\citenamefont {Li}\ \emph {et~al.}(2014)\citenamefont {Li},
  \citenamefont {Xu},\ and\ \citenamefont {Chen}}]{PhysRevB.89.085111}%
  \BibitemOpen
  \bibfield  {author} {\bibinfo {author} {\bibfnamefont {L.}~\bibnamefont
  {Li}}, \bibinfo {author} {\bibfnamefont {Z.}~\bibnamefont {Xu}}, \ and\
  \bibinfo {author} {\bibfnamefont {S.}~\bibnamefont {Chen}},\ }\href {\doibase
  10.1103/PhysRevB.89.085111} {\bibfield  {journal} {\bibinfo  {journal} {Phys.
  Rev. B}\ }\textbf {\bibinfo {volume} {89}},\ \bibinfo {pages} {085111}
  (\bibinfo {year} {2014})}\BibitemShut {NoStop}%
\bibitem [{\citenamefont {P\'erez-Gonz\'alez}\ \emph
  {et~al.}(2019)\citenamefont {P\'erez-Gonz\'alez}, \citenamefont {Bello},
  \citenamefont {G\'omez-Le\'on},\ and\ \citenamefont
  {Platero}}]{PhysRevB.99.035146}%
  \BibitemOpen
  \bibfield  {author} {\bibinfo {author} {\bibfnamefont {B.}~\bibnamefont
  {P\'erez-Gonz\'alez}}, \bibinfo {author} {\bibfnamefont {M.}~\bibnamefont
  {Bello}}, \bibinfo {author} {\bibfnamefont {A.}~\bibnamefont
  {G\'omez-Le\'on}}, \ and\ \bibinfo {author} {\bibfnamefont {G.}~\bibnamefont
  {Platero}},\ }\href {\doibase 10.1103/PhysRevB.99.035146} {\bibfield
  {journal} {\bibinfo  {journal} {Phys. Rev. B}\ }\textbf {\bibinfo {volume}
  {99}},\ \bibinfo {pages} {035146} (\bibinfo {year} {2019})}\BibitemShut
  {NoStop}%
\bibitem [{\citenamefont {Dal~Lago}\ \emph {et~al.}(2015)\citenamefont
  {Dal~Lago}, \citenamefont {Atala},\ and\ \citenamefont
  {Foa~Torres}}]{PhysRevA.92.023624}%
  \BibitemOpen
  \bibfield  {author} {\bibinfo {author} {\bibfnamefont {V.}~\bibnamefont
  {Dal~Lago}}, \bibinfo {author} {\bibfnamefont {M.}~\bibnamefont {Atala}}, \
  and\ \bibinfo {author} {\bibfnamefont {L.~E.~F.}\ \bibnamefont
  {Foa~Torres}},\ }\href {\doibase 10.1103/PhysRevA.92.023624} {\bibfield
  {journal} {\bibinfo  {journal} {Phys. Rev. A}\ }\textbf {\bibinfo {volume}
  {92}},\ \bibinfo {pages} {023624} (\bibinfo {year} {2015})}\BibitemShut
  {NoStop}%
\bibitem [{\citenamefont {Zhu}\ \emph {et~al.}(2018)\citenamefont {Zhu},
  \citenamefont {Zhong}, \citenamefont {Ke}, \citenamefont {Qin}, \citenamefont
  {Sukhorukov}, \citenamefont {Kivshar},\ and\ \citenamefont
  {Lee}}]{PhysRevA.98.013855}%
  \BibitemOpen
  \bibfield  {author} {\bibinfo {author} {\bibfnamefont {B.}~\bibnamefont
  {Zhu}}, \bibinfo {author} {\bibfnamefont {H.}~\bibnamefont {Zhong}}, \bibinfo
  {author} {\bibfnamefont {Y.}~\bibnamefont {Ke}}, \bibinfo {author}
  {\bibfnamefont {X.}~\bibnamefont {Qin}}, \bibinfo {author} {\bibfnamefont
  {A.~A.}\ \bibnamefont {Sukhorukov}}, \bibinfo {author} {\bibfnamefont
  {Y.~S.}\ \bibnamefont {Kivshar}}, \ and\ \bibinfo {author} {\bibfnamefont
  {C.}~\bibnamefont {Lee}},\ }\href {\doibase 10.1103/PhysRevA.98.013855}
  {\bibfield  {journal} {\bibinfo  {journal} {Phys. Rev. A}\ }\textbf {\bibinfo
  {volume} {98}},\ \bibinfo {pages} {013855} (\bibinfo {year}
  {2018})}\BibitemShut {NoStop}%
\bibitem [{\citenamefont {Bergholtz}\ \emph {et~al.}(2021)\citenamefont
  {Bergholtz}, \citenamefont {Budich},\ and\ \citenamefont
  {Kunst}}]{RevModPhys.93.015005}%
  \BibitemOpen
  \bibfield  {author} {\bibinfo {author} {\bibfnamefont {E.~J.}\ \bibnamefont
  {Bergholtz}}, \bibinfo {author} {\bibfnamefont {J.~C.}\ \bibnamefont
  {Budich}}, \ and\ \bibinfo {author} {\bibfnamefont {F.~K.}\ \bibnamefont
  {Kunst}},\ }\href {\doibase 10.1103/RevModPhys.93.015005} {\bibfield
  {journal} {\bibinfo  {journal} {Rev. Mod. Phys.}\ }\textbf {\bibinfo {volume}
  {93}},\ \bibinfo {pages} {015005} (\bibinfo {year} {2021})}\BibitemShut
  {NoStop}%
\bibitem [{\citenamefont {Dobrykh}\ \emph {et~al.}(2018)\citenamefont
  {Dobrykh}, \citenamefont {Yulin}, \citenamefont {Slobozhanyuk}, \citenamefont
  {Poddubny},\ and\ \citenamefont {Kivshar}}]{PhysRevLett.121.163901}%
  \BibitemOpen
  \bibfield  {author} {\bibinfo {author} {\bibfnamefont {D.~A.}\ \bibnamefont
  {Dobrykh}}, \bibinfo {author} {\bibfnamefont {A.~V.}\ \bibnamefont {Yulin}},
  \bibinfo {author} {\bibfnamefont {A.~P.}\ \bibnamefont {Slobozhanyuk}},
  \bibinfo {author} {\bibfnamefont {A.~N.}\ \bibnamefont {Poddubny}}, \ and\
  \bibinfo {author} {\bibfnamefont {Y.~S.}\ \bibnamefont {Kivshar}},\ }\href
  {\doibase 10.1103/PhysRevLett.121.163901} {\bibfield  {journal} {\bibinfo
  {journal} {Phys. Rev. Lett.}\ }\textbf {\bibinfo {volume} {121}},\ \bibinfo
  {pages} {163901} (\bibinfo {year} {2018})}\BibitemShut {NoStop}%
\bibitem [{\citenamefont {Hadad}\ \emph {et~al.}(2016)\citenamefont {Hadad},
  \citenamefont {Khanikaev},\ and\ \citenamefont {Al\`u}}]{PhysRevB.93.155112}%
  \BibitemOpen
  \bibfield  {author} {\bibinfo {author} {\bibfnamefont {Y.}~\bibnamefont
  {Hadad}}, \bibinfo {author} {\bibfnamefont {A.~B.}\ \bibnamefont
  {Khanikaev}}, \ and\ \bibinfo {author} {\bibfnamefont {A.}~\bibnamefont
  {Al\`u}},\ }\href {\doibase 10.1103/PhysRevB.93.155112} {\bibfield  {journal}
  {\bibinfo  {journal} {Phys. Rev. B}\ }\textbf {\bibinfo {volume} {93}},\
  \bibinfo {pages} {155112} (\bibinfo {year} {2016})}\BibitemShut {NoStop}%
\bibitem [{\citenamefont {Boyd}(2020)}]{boyd2020nonlinear}%
  \BibitemOpen
  \bibfield  {author} {\bibinfo {author} {\bibfnamefont {R.~W.}\ \bibnamefont
  {Boyd}},\ }\href@noop {} {\emph {\bibinfo {title} {Nonlinear optics}}}\
  (\bibinfo  {publisher} {Academic press},\ \bibinfo {year} {2020})\BibitemShut
  {NoStop}%
\bibitem [{\citenamefont {Wang}\ \emph {et~al.}(2016)\citenamefont {Wang},
  \citenamefont {Zhang}, \citenamefont {Zhang}, \citenamefont {Luo},
  \citenamefont {Xiong}, \citenamefont {Wang}, \citenamefont {Li},
  \citenamefont {Hu},\ and\ \citenamefont {You}}]{PhysRevB.94.224410}%
  \BibitemOpen
  \bibfield  {author} {\bibinfo {author} {\bibfnamefont {Y.-P.}\ \bibnamefont
  {Wang}}, \bibinfo {author} {\bibfnamefont {G.-Q.}\ \bibnamefont {Zhang}},
  \bibinfo {author} {\bibfnamefont {D.}~\bibnamefont {Zhang}}, \bibinfo
  {author} {\bibfnamefont {X.-Q.}\ \bibnamefont {Luo}}, \bibinfo {author}
  {\bibfnamefont {W.}~\bibnamefont {Xiong}}, \bibinfo {author} {\bibfnamefont
  {S.-P.}\ \bibnamefont {Wang}}, \bibinfo {author} {\bibfnamefont {T.-F.}\
  \bibnamefont {Li}}, \bibinfo {author} {\bibfnamefont {C.-M.}\ \bibnamefont
  {Hu}}, \ and\ \bibinfo {author} {\bibfnamefont {J.~Q.}\ \bibnamefont {You}},\
  }\href {\doibase 10.1103/PhysRevB.94.224410} {\bibfield  {journal} {\bibinfo
  {journal} {Phys. Rev. B}\ }\textbf {\bibinfo {volume} {94}},\ \bibinfo
  {pages} {224410} (\bibinfo {year} {2016})}\BibitemShut {NoStop}%
\bibitem [{\citenamefont {Shen}\ \emph {et~al.}(2021)\citenamefont {Shen},
  \citenamefont {Wang}, \citenamefont {Li}, \citenamefont {Zhu}, \citenamefont
  {Agarwal},\ and\ \citenamefont {You}}]{PhysRevLett.127.183202}%
  \BibitemOpen
  \bibfield  {author} {\bibinfo {author} {\bibfnamefont {R.-C.}\ \bibnamefont
  {Shen}}, \bibinfo {author} {\bibfnamefont {Y.-P.}\ \bibnamefont {Wang}},
  \bibinfo {author} {\bibfnamefont {J.}~\bibnamefont {Li}}, \bibinfo {author}
  {\bibfnamefont {S.-Y.}\ \bibnamefont {Zhu}}, \bibinfo {author} {\bibfnamefont
  {G.~S.}\ \bibnamefont {Agarwal}}, \ and\ \bibinfo {author} {\bibfnamefont
  {J.~Q.}\ \bibnamefont {You}},\ }\href {\doibase
  10.1103/PhysRevLett.127.183202} {\bibfield  {journal} {\bibinfo  {journal}
  {Phys. Rev. Lett.}\ }\textbf {\bibinfo {volume} {127}},\ \bibinfo {pages}
  {183202} (\bibinfo {year} {2021})}\BibitemShut {NoStop}%
\bibitem [{\citenamefont {Nair}\ \emph
  {et~al.}(2021{\natexlab{a}})\citenamefont {Nair}, \citenamefont
  {Mukhopadhyay},\ and\ \citenamefont {Agarwal}}]{PhysRevLett.126.180401}%
  \BibitemOpen
  \bibfield  {author} {\bibinfo {author} {\bibfnamefont {J.~M.~P.}\
  \bibnamefont {Nair}}, \bibinfo {author} {\bibfnamefont {D.}~\bibnamefont
  {Mukhopadhyay}}, \ and\ \bibinfo {author} {\bibfnamefont {G.~S.}\
  \bibnamefont {Agarwal}},\ }\href {\doibase 10.1103/PhysRevLett.126.180401}
  {\bibfield  {journal} {\bibinfo  {journal} {Phys. Rev. Lett.}\ }\textbf
  {\bibinfo {volume} {126}},\ \bibinfo {pages} {180401} (\bibinfo {year}
  {2021}{\natexlab{a}})}\BibitemShut {NoStop}%
\bibitem [{\citenamefont {Yu}\ \emph {et~al.}(2020)\citenamefont {Yu},
  \citenamefont {Shen},\ and\ \citenamefont {Li}}]{PhysRevLett.124.213604}%
  \BibitemOpen
  \bibfield  {author} {\bibinfo {author} {\bibfnamefont {M.}~\bibnamefont
  {Yu}}, \bibinfo {author} {\bibfnamefont {H.}~\bibnamefont {Shen}}, \ and\
  \bibinfo {author} {\bibfnamefont {J.}~\bibnamefont {Li}},\ }\href {\doibase
  10.1103/PhysRevLett.124.213604} {\bibfield  {journal} {\bibinfo  {journal}
  {Phys. Rev. Lett.}\ }\textbf {\bibinfo {volume} {124}},\ \bibinfo {pages}
  {213604} (\bibinfo {year} {2020})}\BibitemShut {NoStop}%
\bibitem [{\citenamefont {Nair}\ \emph
  {et~al.}(2021{\natexlab{b}})\citenamefont {Nair}, \citenamefont
  {Mukhopadhyay},\ and\ \citenamefont {Agarwal}}]{PhysRevB.103.224401}%
  \BibitemOpen
  \bibfield  {author} {\bibinfo {author} {\bibfnamefont {J.~M.~P.}\
  \bibnamefont {Nair}}, \bibinfo {author} {\bibfnamefont {D.}~\bibnamefont
  {Mukhopadhyay}}, \ and\ \bibinfo {author} {\bibfnamefont {G.~S.}\
  \bibnamefont {Agarwal}},\ }\href {\doibase 10.1103/PhysRevB.103.224401}
  {\bibfield  {journal} {\bibinfo  {journal} {Phys. Rev. B}\ }\textbf {\bibinfo
  {volume} {103}},\ \bibinfo {pages} {224401} (\bibinfo {year}
  {2021}{\natexlab{b}})}\BibitemShut {NoStop}%
\bibitem [{\citenamefont {Agarwal}(2012)}]{agarwal2012quantum}%
  \BibitemOpen
  \bibfield  {author} {\bibinfo {author} {\bibfnamefont {G.~S.}\ \bibnamefont
  {Agarwal}},\ }\href@noop {} {\emph {\bibinfo {title} {Quantum optics}}}\
  (\bibinfo  {publisher} {Cambridge University Press},\ \bibinfo {year}
  {2012})\BibitemShut {NoStop}%
\end{thebibliography}%

\end{document}